\documentclass[12pt,a4paper]{article}
\pdfoutput=1
\usepackage{jheppub}

\usepackage{bm}
\usepackage{ifsym}
\usepackage{amsthm}
\usepackage{amsfonts}
\usepackage{ccaption}
\usepackage{mathrsfs}
\usepackage{booktabs}

  \captionnamefont{\bfseries}
  \captiontitlefont{\small\sffamily}
  \captiondelim{: }
  \hangcaption

\def\clock{{\count0=\time
           \divide\count0 60
           \ifnum\count0<10 0\fi\the\count0
           \multiply\count0 -60 \advance\count0 \time
           :\ifnum\count0<10 0\fi \the\count0
         }}
\newcommand{\timestamp}{{\small\vbox{\hbox{\tt\jobname.tex}
\hbox{\the\day/\the\month/\the\year, \clock}}}}

\def\sec#1{\S\ref{#1}}
\def\fig#1{Fig.\,\ref{#1}}
\def\req#1{(\ref{#1})}

\newcommand{\ie}{{\it i.e.,\,}}

\newcommand{\lp}{\left(}
\newcommand{\rp}{\right)}

\newcommand{\beq}{\begin{equation}}
\newcommand{\eeq}{\end{equation}}
\newcommand{\bea}{\begin{eqnarray}}
\newcommand{\eea}{\end{eqnarray}}
\newcommand{\beqa}{\begin{eqnarray}}
\newcommand{\eeqa}{\end{eqnarray}}

\newcommand{\bp}{\mathsf{R_c}}
\newcommand{\bE}{\mathsf{\delta_e}}

\def\sec#1{\S\ref{#1}}
\def\fig#1{Fig.\,\ref{#1}}
\def\req#1{(\ref{#1})}
\def\app#1{Appendix \ref{#1}}

\definecolor{rust}{rgb}{0.8,0.2,0.2}
\definecolor{green}{rgb}{0.1,0.8,0.2}

\numberwithin{equation}{section}

\title{Effective hydrodynamics of black D3-branes}
\author{Roberto Emparan$^{a,b}$, }
\author{Veronika E. Hubeny$^c$}
\author{ \&  Mukund Rangamani$^c$}

\affiliation[a]{Instituci\'o Catalana de Recerca i Estudis
Avan\c cats (ICREA) \\ 
Passeig Llu\'{\i}s Companys 23, E-08010 Barcelona, Spain.}
\affiliation[b]{Departament de F{\'\i}sica Fonamental and \\
Institut de Ci\`encies del Cosmos, Universitat de Barcelona, \\
Mart\'{\i} i Franqu\`es 1, E-08028 Barcelona, Spain.}
\affiliation[c]{ Centre for Particle Theory \& Department of Mathematical Sciences,\\
Science Laboratories, South Road, Durham DH1 3LE, UK.}

\emailAdd{emparan@ub.edu}
\emailAdd{veronika.hubeny@durham.ac.uk}
\emailAdd{mukund.rangamani@durham.ac.uk}

\abstract{
The long-wavelength effective field theory of world-volume fluctuations of black D3-branes is shown to be a hydrodynamical system to leading order in a gradient expansion.
We study the system on a fiducial `cutoff' surface: the fluctuating geometry imprints its dynamics on the surface via an induced stress tensor whose conservation encapsulates the hydrodynamical description.
For a generic non-extremal D3-brane, as we move our cutoff surface from the asymptotically flat near-boundary region to the near-horizon region,
this hydrodynamical system interpolates between a non-conformal relativistic fluid and a non-relativistic incompressible fluid.  We also consider the dependence on the deviation from extremality of the D3-branes.
In the near-extremal case we recover the description in terms of a conformal relativistic fluid encountered in the AdS/CFT context. We argue that this system allows us therefore to explore the various connections that have hitherto been suggested relating the dynamics of gravitational systems and fluid dynamics. In particular, we go on to show that the blackfold effective field theory approach allows us to capture this hydrodynamical behaviour and moreover subsumes the constructions encountered in the fluid/gravity correspondence and the black hole membrane paradigm, providing thereby a universal language to explore the effective dynamics of black branes. } 


\begin{document}
\begin{flushright} \small{DCPT-13/11} \end{flushright}

\maketitle

\flushbottom

\section{Introduction}
\label{s:intro}

Black holes and black branes, which are solutions to gravitational field equations, are generically expected to have complex  (classical) dynamics. There are however certain useful regimes where the non-linear dynamics governing their behaviour can be mapped to a more familiar classical dynamical system, viz., hydrodynamics. General relativists will undoubtedly be familiar with the black hole membrane paradigm \cite{Damour:1978cg,Damour:1982fk, Price:1986yy} which identifies the dynamics of 
the black hole horizon as being that of a non-relativistic fluid. More recently, the low energy dynamics of planar black holes in asymptotically AdS spacetimes was shown to be given by that of a relativistic conformal fluid \cite{Bhattacharyya:2008jc}, a relation that has come to be known as the fluid/gravity correspondence \cite{Hubeny:2011hd}. In a parallel development, attempts to understand the nature of black holes in more than four dimensions have led to the development of the `blackfold' paradigm \cite{Emparan:2009cs,Emparan:2009at,Camps:2010br,Camps:2012hw}, which identifies the low energy effective field theory of black branes. A sector of this effective description takes the form of a hydrodynamic theory.

It is obvious that we suffer from an embarrassment of riches when it comes to identifying the effective dynamics of black holes. We seem to have zeroed in on a set of low energy degrees of freedom which obey hydrodynamic equations. However,  in each of the cases described above, these degrees of freedom live in different parts of the spacetime:  (a) the  membrane paradigm would have them reside on the horizon (or the stretched horizon),  (b) the fluid/gravity correspondence takes them all the way to the asymptotic boundary of the spacetime, and (c) the blackfold approach identifies them as living in an overlap  zone, which is the transitory zone between the near horizon and the asymptotic region.  Suffice it to say that such a situation calls for a clearer understanding of how these various constructions are related to each other. 

The main aim of the current work is to provide a simple model where we can study the relation between these constructions explicitly and identify how the hydrodynamical degrees of freedom morph across the various descriptions. In particular, we will describe a simple inclusion relation 
\begin{equation}
\text{Membrane paradigm } \subset \text{Fluid/gravity correspondence }\subset \text{Blackfolds}  \,.
\label{inclusionrel}
\end{equation}	
While such a relation ought to have been anticipated (it is partly clear from our description above), by picking a specific example to illustrate this inclusion, we will be able to track the change in the hydrodynamic description via the variation of the constitutive relations (and thereby the transport coefficients). 

Before we proceed to describe the particular set-up we use, we should emphasize that we will be a bit fast and loose with the notion of the membrane paradigm in what follows. Rather than working with the traditional description derived in \cite{Damour:1982fk, Price:1986yy}, we will use a modern interpretation of the same in terms of an effective Rindler fluid \cite{Bredberg:2011jq}. The claim is the following: since the near-horizon geometries of non-extremal black branes is universally a Rindler geometry, by analyzing the dynamics of gravity in this Rindler patch, one can infer the dynamics of the stretched horizon relevant for the membrane paradigm. An explicit analysis following the long-wavelength perturbation theory analogous to that employed in the fluid/gravity correspondence, together with Dirichlet boundary conditions at the end of the Rindler universe, leads to a map between the gravitational degrees of freedom and a non-relativistic incompressible fluid. These ideas have been extensively explored in the recent literature; see \cite{Compere:2011dx, Compere:2012mt, Eling:2012ni} for further developments showing close connection with the fluid/gravity correspondence.\footnote{
Earlier analysis of the Rindler near-horizon behaviour in the linearized regime which motivated the Rindler fluid correspondence was considered in \cite{Bredberg:2010ky} and further analyzed in detail in \cite{Marolf:2012dr}. For a selection of references that explore these considerations in a varied class of gravitational models see \cite{Cai:2011xv,Kuperstein:2011fn,Cai:2012vr,Matsuo:2012pi,Bai:2012ci,Cai:2012mg}. A discussion of the membrane paradigm from other viewpoints motivated in part by the fluid-gravity correspondence can be found in \cite{Eling:2009pb,Eling:2009sj,Eling:2011ct,Brattan:2011my}.} A critical discussion of the connections between this approach and the  conventional story will appear elsewhere \cite{Emparan:2013fk}. 

The basic picture that we will flesh  out in detail is the following: we start with the non-extremal black D3-brane geometry, which is an asymptotically flat solution to Type IIB supergravity equations of motion. As is well known, in the regime where the solution is near-extremal the geometry develops an $\text{AdS}_5 \times {\bf S}^5$ throat and one can view the near-extremal solution as a planar black hole in AdS spacetime (homogeneously smeared over the ${\bf S}^5$). Zooming in further towards the horizon, we encounter a Rindler region (locally ${\mathbb R}^{4,1} \times {\bf S}^5$). Since the entire geometry encompasses three distinct regions -- the asymptotic ${\mathbb R}^{9,1}$, the AdS throat, and the ultra near-horizon Rindler geometry -- we can investigate the behaviour of the low energy effective description by studying the fluid dynamical system that encodes (in a dual sense) the dynamics of the branes in these three domains. 

To isolate the effective hydrodynamical degrees of freedom, we will enclose the D3-branes in a box and impose Dirichlet boundary conditions on the walls of the box for the dynamical fields including gravity. With these boundary conditions it is a straightforward exercise to allow long-wavelength world-volume fluctuations along the D3-branes and use the supergravity equations of motion to identify the dual hydrodynamical system. The construction closely follows the original discussions in the fluid/gravity context \cite{Bhattacharyya:2008jc} and the 
study of the hydrodynamic description of neutral black branes  in the blackfold approach \cite{Camps:2010br}.\footnote{The effective dynamics of D$p$-branes at perfect fluid level has been described (although not derived from the supergravity equations) in \cite{Emparan:2011hg}.
The hydrodynamic description of neutral branes has been recently extended to include higher order non-linear effects in \cite{Caldarelli:2012hy}.} We only need include the effects of the Dirichlet boundary condition; this  has been previously described in \cite{Brattan:2011my}  and  \cite{Emparan:2012be} in the context of asymptotically AdS and asymptotically flat spacetimes, respectively. We will find that by a suitable combination of the above set of ideas one can easily recover the effective field theory of the black D3-branes.

The upshot of our analysis is an explicit determination of the constitutive relations for the D3-brane world-volume fluctuations in the long-wavelength limit. We will show that this is given by a hydrodynamical system from which we can read off the transport coefficients. Since we are interested in the dynamics on a fixed stack of D3-branes, we will fix the total charge, allowing only the overall energy density to fluctuate. As a consequence, we will find a neutral relativistic fluid capturing the low energy modes on surfaces outside the horizon. Since the spacetime has two non-trivial scales, viz., the horizon size and the cutoff scale measured say in terms of the charge radius of the black brane, we see that by taking different limits we can recover the three descriptions mentioned in \eqref{inclusionrel}.  This neutral fluid tends towards a non-relativistic Rindler fluid in the near-horizon limit as one expects on general grounds following \cite{Bredberg:2011jq}. In the throat region of near-extremal branes, where we recover the AdS geometry, we encounter a conformal fluid along the lines of the fluid gravity correspondence. However, the general description that is valid for all choices of the scales is the blackfold fluid. 
In short, the D3-brane geometry constructs for us a convenient gravitational background which interpolates between multiple regions: Rindler, AdS throat, and asymptotically flat spacetimes. By suitably choosing to place the stretched horizon (or loosely speaking `holographic screen') in the different domains, we get to explore the different constructions encountered earlier in the literature as limiting behaviour of a universal long-wavelength dynamics of black branes.

The outline of this paper is as follows: we begin in \sec{s:intrinsic} with the basic set-up of the D3-brane geometry and collect the necessary ingredients for our computation. In \sec{s:findir} we describe the approximate solution to the equations of motion obtained in a long-wavelength perturbation expansion. The analysis unfortunately is rather technical owing to the complexity of the seed black D3-brane geometry, though the final result giving the constitutive relations for the fluid dynamical system is quite simple and intuitive. We describe some of the salient features of the hydrodynamics in \sec{s:physics} -- the cursory reader may find it useful to focus on this section to get an overall picture before diving into the nitty-gritty details of the preceding section. We conclude with a brief discussion of the salient points in \sec{s:discuss}. The appendices collect some technical results relating to the dimensional reduction and the construction of counter-terms which we use in the main part of the analysis.

\section{Preliminaries: The D3-brane geometry and boundary conditions}
\label{s:intrinsic}

We are interested in studying the intrinsic hydrodynamic behaviour of D3-branes, wherein we allow arbitrary long-wavelength fluctuations along the world-volume directions. Let us begin by recalling the D3-brane solution of Type IIB supergravity \cite{Horowitz:1991cd}
\begin{equation}
ds^2 = -\Delta_+ \,
\Delta_-^{-\frac{1}{2}}\, dt^2 +
\Delta_-^\frac{1}{2}\, d{\bf x}^2 + \frac{dr^2}{\Delta_+\,
\Delta_-} +r^2 \,d\Omega_5^2
\label{hsD3}
\end{equation}	
where
\begin{equation}
\Delta_\pm = 1-\frac{r_\pm^4}{r^4} \ .
\label{deltapm}
\end{equation}	
This solution is supported by the self-dual  5-form field strength, which is given by 
\begin{equation}
F_{(5)}= Q\left(1+\star\right) \text{Vol}({\bf S}^5)\,.
\label{}
\end{equation}	
The parameter $Q$ is related to the horizon radii of the black D3-brane solution
\begin{equation}
Q=2 \, r_+^2 \, r_-^2
\label{Qrprm}
\end{equation}	
and is, up to a conventional factor, the D3 charge 
\begin{equation}
\mathbf{Q}=\frac{\Omega_5}{8\pi \,G_{N}^{(10)}}\; Q \equiv \frac{1}{\kappa_5^2} \, Q\,.
\label{d3charge}
\end{equation}	
where we introduced a normalization factor $\kappa_5$ which will prove useful in the sequel.\footnote{Note that as defined $\kappa_5^2$ has an unconventional length dimension of $8$.} It is the effective five dimensional gravitational coupling for the KK reduced theory which we will introduce shortly.

Since we are interested in the intrinsic dynamics i.e., dynamics along the D3-brane world-volume, it is useful to rewrite the solution in an explicit world-volume covariant form
\begin{equation}
ds^2 =
\Delta_-^{\frac{1}{2}}\left( -\frac{\Delta_+}{\Delta_-}u_a u_b+P_{ab}\right)d\sigma^a d\sigma^b + \frac{dr^2}{\Delta_+\,
\Delta_-} +r^2 \, d\Omega_5^2\,.
\end{equation}	
We have introduced a world-volume (normalized) 4-velocity parameter $u^a$ to achieve world-volume covariance ($P_{ab} = h_{ab} + u_a\, u_b$) and thus extended the static solution of the black D3-brane to a stationary one.\footnote{
The world-volume coordinates will be henceforth denoted as $\sigma^a = \{ t, {\bf x}\} $ and we will use lower-case Latin letters to describe these directions; these indices are  raised and lowered with the world-volume metric $h_{ab}$ which for the most part will be taken to be the flat Minkowski metric $h_{ab} = \eta_{ab}$. Lower-case Greek indices will denote five dimensional bulk directions (with metric $g_{\mu\nu}$) and when necessary we will use upper-case Latin indices to indicate the full ten dimensional coordinates (with metric $G_{AB}$).} This makes it manifest that we have at hand a five parameter family of solutions of Type IIB supergravity: 2 of the parameters are the mass and charge encoded in the horizon radii $r_\pm$ and the remaining three are the horizon velocities captured by $u^a$.\footnote{
We could also consider spinning up the branes in 10 dimensions by turning on rotation along the ${\bf S}^5$ directions. This is captured by three more parameters, the angular momentum chemical potentials residing in the Cartan of $SO(6)$. These should be easy to include in our analysis (though we don't for the sake of simplicity) and would give rise to the dynamics of a charged fluid carrying $U(1)^3$ charges.}

We are interested in dynamics that keeps $Q$ fixed and varies $r_+$ and $u^a$, whilst leaving undeformed the shape of the ${\bf S}^5$; the radius of this transverse sphere will however change from position to position along the world-volume. It therefore turns out to be convenient to  perform a KK reduction along the ${\bf S}^5$ and use the breathing mode scalar as a dynamical field in what follows.

\subsection{The five dimensional effective action}
\label{s:5deff}

The KK reduction of the fields on the ${\bf S}^5$ can be achieved by standard techniques. We start with the 10D fields written in a warped product from 
\begin{equation}
ds^2=ds^2_5+e^{2\,\varphi}\, d\Omega_5^2 \,, \qquad F_{(5)}= Q\left( e^{-5\,\varphi}\, \epsilon_{(5)} +
\text{Vol}({\bf S}^5)\right)\,.
\label{}
\end{equation}	
As we fix the charge it transpires that the reduction to the five dimensional metric and the dilaton $\varphi$ is a consistent truncation that has been discussed previously in the literature \cite{Bremer:1998zp, Liu:2000gk}. 
In any event it is a simple matter to obtain the reduced action as described in \app{s:kkreduce},
\begin{equation}
I_5= \frac{1}{2\, \kappa_5^2}
\int d^5 x\; \sqrt{-g}\,e^{5\,\varphi}\left(R+20\, (\partial\varphi)^2+20\, e^{-2\,\varphi} -2\, Q^2\, e^{-10\,\varphi}\right) .
\label{5deffn}
\end{equation}	
This effective action needs to be supplemented with appropriate boundary terms (and possibly counter-terms), which we will address in due course.  The effective 5D equations arising from \eqref{5deffn} are 
\begin{align}
\text{Eom}_{\mu\nu}: & \qquad R_{\mu\nu} =5\, (\nabla_\mu\varphi\, \nabla_\nu\varphi+\nabla_\mu\nabla_\nu\varphi)-Q^2e^{-10\varphi}\; g_{\mu\nu} 
\nonumber \\
\text{Eom}_\varphi : & \qquad \Box\varphi+5(\partial\varphi)^2 =4\,e^{-2\varphi}-Q^2\,e^{-10\,\varphi}\,.
\label{Eommunuphi}
\end{align}
The metric and dilaton of the static D3-brane solution are immediately read off as
\begin{equation}
ds_5^2 = -\Delta_+ \,
\Delta_-^{-\frac{1}{2}}\, dt^2 +
\Delta_-^\frac{1}{2}\, d{\bf x}^2 + \frac{dr^2}{\Delta_+\,
\Delta_-} \ ,\qquad
e^{\varphi} = r\,.
\label{5dsoln}
\end{equation}	
and it is obvious how to pass over to the world-volume covariant form.

\subsection{The world-volume stress tensor}
\label{s:5deff}

Having obtained the effective five dimensional system of equations we now proceed to describe the construction of the world-volume stress energy-momentum tensor. As mentioned in \sec{s:intro}, we  are interested in enclosing the brane in a box of radius $R$. The boundary conditions we want to impose involve fixing the world-volume metric $h_{ab}$ and the dilaton $\varphi$ on this surface. In fact, since we have a scalar field we can adapt our coordinate chart to level sets of the scalar and declare the hypersurface where we impose the boundary condition to be given by $\varphi = \text{constant}$. 

This choice of boundary conditions requires the addition of appropriate boundary terms to the action $I_5$ \eqref{5deffn}. We will also include some counter-terms. Whilst strictly not necessary to ensure the finiteness of the on-shell action (as long as the D3-brane is enclosed within a box of finite radius $r=R$), they will prove useful in the study of various limits, since the counter-terms perform a subtraction of stress-energies from ground states that do not affect the hydrodynamic behaviour. For instance, we would like to recover the conformal fluid dynamics when we enter the throat region and this allows us to isolate an appropriate set of counter-terms. We also assume that $\varphi$  is constant along the boundary, i.e., $h^{ab}\, \nabla_b \varphi=0$ and furthermore, to leading order in our analysis we take  $h_{ab}$ to be a flat metric. This is an important simplifying assumption, which in particular allows us to easily embed the boundary $\varphi = \text{constant}$ in a $D=5$ dimensional Minkowski space (thereby making it easy to infer the counter-terms by a background subtraction scheme). 

We claim that the complete $5$-dimensional action, with all Gibbons-Hawking terms and counter-terms is (a detailed derivation is presented in \app{s:qlocalst})
\begin{equation}
I_{5\text{tot}}=I_5+I_{GH}+I_{GH\varphi}+I_{ct}\,.
\label{I5tot}
\end{equation}
These counter-terms we present below are sufficient for $h_{ab} = \eta_{ab}$ being the flat metric and $\varphi$ held constant along the world-volume. If we wish to let these fields vary, we would need to include additional counter-terms involving the world-volume intrinsic curvature and the gradient of $\varphi$.

The terms with derivatives normal to the surface can be derived from the ten-dimensional Gibbons-Hawking term. We split them conveniently in terms of one boundary term for the metric:
\begin{equation}
I_{GH}=  \frac{1}{\kappa_5^2} \; 
\int_{\partial\mathcal{M}} d^{4} x\;\sqrt{-h}\;e^{5 \,\varphi}\;\Theta\,,
\label{}
\end{equation}	
and the other for the dilaton
\begin{equation}
I_{GH\varphi}=\frac{1}{2\, \kappa_5^2} \; \int_{\partial\mathcal{M}} d^{4}x\;\sqrt{-h} \; n^\mu \partial_\mu e^{5\, \varphi}\,.
\label{GHphi}
\end{equation}	
Here  $\Theta$ is the trace of the five-dimensional extrinsic curvature tensor. 

The counter-term action $I_{ct}$ is obtained by removing from the action a constant piece proportional to the non-fluctuating D3-brane charge in addition to the contribution of the ${\bf S}^5$ curvature.  These contributions are explicitly given as 
\begin{equation}
I_{ct}=\frac{1}{2\, \kappa_5^2}\; \int_{\partial\mathcal{M}} d^{4}x\;\sqrt{-h} \left(Q-5\, e^{4\, {\boldsymbol\varphi}}\right)\,.
\label{}
\end{equation}	

The above expressions can be massaged further: since we are assuming that the boundary can be defined as an isodilatonic surface, the unit-normal is
\begin{equation}
n_\mu=\frac{\partial_\mu{\varphi}}{\sqrt{(\partial_n{\varphi})^2}}\,.
\label{}
\end{equation}	
In writing $\partial_n{\varphi}$ we are emphasizing that the derivative is non-zero only along normal directions. Our conventions are that $\varphi$ grows outwards from the surface so that $n^\mu$ points outwards. This can be used to write the above expressions more compactly. In particular, the conserved quasi-local stress tensor induced on the boundary $\varphi = \text{constant}$ (obtained by varying the boundary and counter-term actions with respect to the boundary metric $h_{ab}$) is given by:
\begin{equation}
\label{Tab2}
T_{ab}=\frac{1}{ \kappa_5^2} \;  \left[e^{5\,{\varphi}}\lp\Theta_{ab}-h_{ab} \Theta\rp +\left(5\, e^{5\, {\varphi}} (\sqrt{(\partial_n{\varphi})^2}-e^{-{\varphi}})+Q\right)h_{ab} \right]\,.
\end{equation}
Naively it might seem that we obtain a non-local term for the dilaton. However, this is not the case, as this is a Gibbons-Hawking-type term with normal derivatives, not world-volume ones. More importantly,  this term does not descend from any non-local ten-dimensional counter-term; no such is necessary for the variational problem at hand.

\section{The black D3-brane Dirichlet problem at finite $R$}
\label{s:findir}

We have now gathered all the necessary ingredients to study the dynamics of fluctuating D3-branes with the induced metric and size of the transverse ${\bf S}^5$ held rigid on a timelike hypersurface.  Our starting point is the five dimensional system of fields which are given by \eqref{5dsoln} and we solve the equations of motion \eqref{Eommunuphi} arising from \eqref{5deffn}. 

\subsection{The seed metric for long-wavelength fluctuations}
\label{s:zeroD3}

As indicated earlier the surface where we impose boundary condition is going to be defined relationally as the level set of the dilaton field $\varphi$. It is useful to gauge fix the five dimensional solution so that $e^\varphi = r$ and thus we will impose Dirichlet boundary conditions on the hypersurface $r = R$ for the fields involved.  This takes care of one of the boundary conditions we will be imposing on our fluctuations. The second boundary condition will simply demand regularity of the solution in the interior. As in the fluid/gravity correspondence, we will require that the solution we study in the long-wavelength perturbation expansion have a regular event horizon.

To make issues relating to regularity clear we start with a manifestly regular metric for our perturbation expansion.  Let us first define an ingoing coordinate $v$ via:
\begin{equation}
v = t + \int dr' \frac{1}{\Delta_+\,\Delta_-^\frac{1}{4}} \equiv t + r_\ast
\label{vingo}
\end{equation}	
so as to bring the metric \eqref{5dsoln} into the ingoing Eddington-Finkelstein form:
\begin{equation}
ds^2_5 = 2\, \Delta_-^{-\frac{3}{4}}\, dv\,dr - \Delta_+\, \Delta_-^{-\frac{1}{2}} \, dv^2 + \Delta_-^{\frac{1}{2}} \, d{\bf x}^2\,.
\label{}
\end{equation}	
Introducing the velocity field $u^a$ and letting $\{ v, {\bf x} \} \equiv \sigma^a$ we have the above metric to be written in a familiar form:
\begin{align}
ds^2_5 = - 2\, \Delta_-^{-\frac{3}{4}} \, u_a\, d\sigma^a\,dr - \Delta_+\, \Delta_-^{-\frac{1}{2}}  \, u_a\, u_b\, d\sigma^a\, d\sigma^b  + \Delta_-^{\frac{1}{2}}\, P_{ab}\, d\sigma^a\, d\sigma^b\,.
 \label{zerog1}
\end{align}	
This is almost the seed metric for our perturbation analysis.

Before we proceed let us note that there is one more piece of simplification we can make. We want to hold the D3-brane charge fixed -- we therefore work in terms of the variables $\ell_Q$ and $r_+$, with the charge length $\ell_Q$ defined as
\begin{equation}
r_- =    \frac{\ell_Q^2}{r_+} \ , \qquad \ell_Q^2  \equiv \sqrt{\frac{Q}{2}}
\label{lqdef}
\end{equation}	
to keep formulae simple.

\subsection{Long-wavelength fluctuations \& intrinsic dynamics}
\label{s:}

So far we have described a family of stationary D3-brane solutions. We are however interested in scenarios where the dynamics on the world-volume of the D3-brane is non-trivial. To this end we would like to let the fields vary along the world-volume directions. Intuition from the fluid/gravity correspondence and the blackfold paradigm would indicate that we should anticipate that such dynamics should be given by that of an effective fluid dynamics. Our task is to identify the constitutive relations for such a fluid which should be obtained directly from the supergravity field equations. 

Following thus the standard logic of the fluid/gravity and blackfold approaches, we start by promoting the parameters 
 $u^a$ and $r_+$ (recall $r_-$ has been eliminated in terms of $r_+$ and $\ell_Q$ with the latter held fixed) to functions of $\sigma^a$. This will be the seed metric for our perturbation analysis. We need to find corrections to the this metric that continue to solve the bulk field equations order by order in a world-volume gradient expansion. To wit, the five dimensional geometry is given as 
\begin{equation}
ds^2 = g^{(0)}_{\mu\nu} \, dx^\mu\, dx^\nu  + \sum_{k=1}^\infty \, \epsilon^k \, g^{(k)}_{\mu\nu} \, dx^\mu\, dx^\nu
\label{}
\end{equation}	
with $g^{(k)}_{\mu\nu}$ being $k^{\rm th}$ order in the $\sigma^a$ derivatives. We will determine $g^{(1)}_{\mu\nu}$ explicitly in the sequel.

Before we proceed let us note one issue that necessitates a small change in our seed metric. Recall that we want to set up our problem as a Dirichlet problem with the boundary condition defined relationally by demanding fields take a fixed value on the surface with $e^\varphi = R$. When we vary the bulk metric by promoting $u^a$ and $r_+$ to fields dependent on $\sigma^a$ we do have to ensure that this is satisfied at each order we are working in. At zeroth order, the natural induced metric on the surface $r = R$ from the bulk metric \eqref{zerog1} is not the canonical Minkowski metric in the coordinates used there. We can however bring this to the standard form by a simple rescaling of the coordinates. Consider then as the seed metric 
\begin{equation}
ds^2_{5,(0)} = - 2\, \frac{\Delta_-^{-\frac{3}{4}}}{\Delta_{-R}^{-\frac{1}{4}}\, \Delta_{+ R}^{\frac{1}{2}} } \, u_a\, d\sigma^a\,dr - \frac{\Delta_+\, \Delta_-^{-\frac{1}{2}}}{\Delta_{+R}\, \Delta_{-R}^{-\frac{1}{2}}}
 \, u_a\, u_b\, d\sigma^a\, d\sigma^b 
+ \frac{\Delta_-^{\frac{1}{2}}}{\Delta_{-R}^{\frac{1}{2}} } \, P_{ab}\, d\sigma^a\, d\sigma^b 
 \label{5dseed}
\end{equation}	
where to keep equations compact we introduced a notation for the functions evaluated at the cutoff surface
\begin{equation}
\Delta_{\pm R} = \Delta_{\pm} (R)\,.
\label{}
\end{equation}	

The advantage of working with the rescaled coordinates is that it is easy to keep track of all the effects of the spatio-temporal variations of the fields $u^a$ and $r_+$. Essentially since the functions $\Delta_\pm$ depend on $r_+$ when we allow the geometry to fluctuate, the brane has to contort appropriately to ensure that it respects the box it is enclosed in. These contortions will imprint themselves on the classical dynamics (for instance in terms of stability) and thermodynamics. For neutral branes this has been previously explained in \cite{Emparan:2012be}.

Many of the expressions we encounter in intermediate computations involve various complicated functions. To keep the notation somewhat compact, it is convenient to introduce variables to keep track of the position of the cutoff surface and the deviation from extremality. We therefore define
\begin{equation}
\bp \equiv \Delta_{+R} \,, \qquad \bE \equiv \Delta_-(r_+) \,.
\label{pEpars}
\end{equation}	
$\bp  \in [0,1]$ characterizes positioning of the surface  ($\bp  =0$ corresponds to $R=r_+$, whereas $\bp =1$ corresponds to $R=\infty$), and  $\bE \in [0,1]$ characterizes deviation from extremality  ($\bE=0$ corresponds to the extremal brane $\ell_Q=r_+$, whereas $\bE=1$ corresponds to the neutral brane $\ell_Q=0$). 

Thus the most convenient variables to parametrize our system are $\ell_Q$, $\bp$ and $\bE$. In terms of them, the parameters that we used initially are
\begin{equation}
r_\pm =\ell_Q (1-\bE)^{\mp 1/8}\,,\qquad R=\frac{\ell_Q}{(1-\bE)^{1/8}(1-\bp)^{1/4}}\,. 
\label{rpmRbEbp}
\end{equation}
We will often express our results in these variables to extract the limiting behaviour of the black D3-brane dynamics in the long-wavelength limit in various regimes.

\subsection{Details of the perturbation analysis at first order}
\label{s:d3pert}

To carry out the explicit construction of the fluctuating D3-brane, let us introduce a local basis of operators to work in. Define:
\begin{equation}
u_a\, d\sigma^a = -dv + \beta_i(v,x)\, dx^i \,.
\label{}
\end{equation}	
We will work in a local inertial frame, so that $\beta_i$ vanishes at the leading order. It will give a contribution to the first order equations of motion.  With this understanding the leading order changes in the metric $g^{(0)}_{\mu\nu}$ are given by: 
\begin{align}
- 2\, \frac{\Delta_-^{-\frac{3}{4}}}{\Delta_{-R}^{-\frac{1}{4}}\, \Delta_{+R}^{\frac{1}{2}} } \, u_a\, d\sigma^a\,dr &=  2\,\frac{\Delta_-^{-\frac{3}{4}}}{\Delta_{-R}^{-\frac{1}{4}}\, \Delta_{+R}^{\frac{1}{2}} }  \left[dv\,dr - \beta_i\, dx^i\, dr\right] +2 \frac{\delta}{\delta r_+}
\left(\frac{\Delta_-^{-\frac{3}{4}}}{\Delta_{-R}^{-\frac{1}{4}}\, \Delta_{+ R}^{\frac{1}{2}} } \right)
 \, \delta r_+ \;dr \, dv  \,,
\nonumber \\
- \frac{\Delta_+\, \Delta_-^{-\frac{1}{2}}}{\Delta_{+R}\, \Delta_{-R}^{-\frac{1}{2}}} \, u_a\, u_b\, d\sigma^a\, d\sigma^b  &= 
-  \frac{\Delta_+\, \Delta_-^{-\frac{1}{2}}}{\Delta_{+R}\, \Delta_{-R}^{-\frac{1}{2}}}  \left[dv^2 -2\, \beta_i\, dx^i\, dv\right] - \frac{\delta}{\delta r_+} \left( \frac{\Delta_+\, \Delta_-^{-\frac{1}{2}}}{\Delta_{+R}\, \Delta_{-R}^{-\frac{1}{2}}}  \right) \delta r_+\, dv^2 \,,
\nonumber \\
\frac{\Delta_-^{\frac{1}{2}}}{\Delta_{-R}^{\frac{1}{2}} } \, P_{ab}\, d\sigma^a\, d\sigma^b &= \frac{\Delta_-^{\frac{1}{2}}}{\Delta_{-R}^{\frac{1}{2}} }  \, \left[d{\bf x}^2 -2\, \beta_i\, dx^i\, dv\right]+ \frac{\delta}{\delta r_+} \left(\frac{\Delta_-^{\frac{1}{2}}}{\Delta_{-R}^{\frac{1}{2}} } \right) \delta r_+\, d{\bf x}^2 \,.
\end{align}

We are now in a position to solve the equations of motion arising from \eqref{5deffn} to ascertain the correction to the metric at first order $g^{(1)}_{\mu\nu}$. Since the background solution of the static D3-brane \eqref{5dsoln} is $SO(3)$ symmetric, it is useful to exploit this spatial symmetry. We therefore look for fluctuations in the different symmetry sectors, which are guaranteed to remain decoupled in our perturbation expansion at each order. Since this construction follows the standard logic of the fluid/gravity correspondence we will be sketchy in presenting all the details of the derivation, preferring instead to dwell on the parts of the construction that are somewhat novel to the case at hand (mostly owing to the Dirichlet boundary conditions). We present the salient results of each of the symmetry sectors in turn.

\subsubsection{The tensors of $SO(3)$}
\label{s:fintensor}

The tensor sector is the easiest since there are no constraints that we need to keep track of and the dilaton plays a passive role (it is a scalar under $SO(3)$). Parameterizing the first order correction as 
\begin{equation}
ds^2_{(1),T} = \,\frac{\Delta_-^{\frac{1}{2}}}{\Delta_{-R}^{\frac{1}{2}} } \,\alpha_{ij}(r) \, dx^i\, dx^j \,,
\label{astenpar}
\end{equation}	
we have a simple equation of motion for the fluctuation field $\alpha_{ij}(r)$:
\begin{equation}
\frac{d}{dr}\left(r^5\, \Delta_+\, \Delta_- \, \frac{d\alpha_{ij}}{dr}\right) = -\,2\; \frac{\Delta_{+R}^\frac{1}{2}}{\Delta_{-R}^{\frac{1}{4}}} \; \left[\Delta_-^{-\frac{1}{4}} \left(5\,r^4 - \frac{2\, \ell_Q^8}{r_+^4}\right) \right] \sigma_{ij} \,.
\label{}
\end{equation}	
This is the minimally coupled scalar wave equation in the background \eqref{5dsoln} with a source set by the shear tensor $\sigma_{ij} = \partial_{(i} \beta_{j)} - \frac{1}{3}\, \delta_{ij}\, \partial_k \, \beta^k$.

The differential operator is invertible and moreover the source in the r.h.s.\ is an exact differential which can be integrated directly. There are two constants of integration: one of these is fixed by our Dirichlet boundary condition and the second by demanding that the solution be regular in the interior. Operationally the latter is fixed by demanding that there is no singularity in $\alpha_{ij}$ as $r \to r_+$. We find:
\begin{equation}
r^5\, \Delta_+\, \Delta_- \, \frac{d\alpha_{ij}}{dr} = c_{ij} -2 \, \left( \frac{\Delta_{+R}^{\frac{1}{2}}}{\Delta_{-R}^{\frac{1}{4}}}\right) r^5\, \Delta_-^{\frac{3}{4}} \;\sigma_{ij}\,.
\label{}
\end{equation}	
One of  the boundary conditions we want to impose is the regularity of the solution at the putative horizon.  This can be achieved by choosing the constant of integration $c_{ij}$  appropriately. Essential owing to the $\Delta_+$ in the kinetic term we demand that the r.h.s. in the above expression has a simple zero at $r  =r_+$. This gives:
\begin{align}
\alpha_{ij}(r)  &=-2\, \sigma_{ij}\,\frac{\Delta_{+R}^{\frac{1}{2}}}{\Delta_{-R}^{\frac{1}{4}}} \left[(R_\ast- r_\ast) - \frac{1}{4}\; \frac{r_+^5\; \Delta_{-R}^{\frac{3}{4}}}{r_+^4- r_-^4} \, \log\left( \frac{\Delta_{+R}}{\Delta_{-R}} \;  \frac{\Delta_-}{\Delta_+} \right) \right] \\
& = - 2\,\sigma_{ij}\, \frac{1}{\Delta_{+R}^{\frac{1}{2}}\,\Delta_{-R}^{\frac{5}{4}}} \; \left(\Delta_{-R}^{\frac{3}{4}}  - \frac{r_+^5}{R^5} \, \Delta_{-}(r_+)^{\frac{3}{4}}\right) (r-R) + \cdots
\label{alphaijR}
\end{align}	
where $r_*$ is the tortoise coordinate defined in \eqref{vingo} and $R_*$ is its value at $r= R$. For convenience we have also written the leading correction at the surface $r = R$ which is essentially all that one needs for computing the world-volume stress tensor. For later convenience we write the solution of tensor perturbations in terms of a single function ${\mathbf t}$,
\begin{equation}
{\mathbf t}(r,\sigma^a) \,\sigma_{ij} \equiv \frac{\Delta_-^{\frac{1}{2}}}{\Delta_{-R}^\frac{1}{2}}\, \alpha_{ij} \,.
\label{tfn}
\end{equation}	
%

\subsubsection{The vector of $SO(3)$}
\label{s:finvector}

We now turn our attention to the vector sector. There are a-priori two sources of metric corrections at this order, since both $g_{vi}$ and $g_{ri}$ transform under the spatial $SO(3)$ rotations as vectors. Hence we shall parameterize our first order metric correction as:
\begin{equation}
ds^2_{(1),V} = 2 \left(\frac{\Delta_+\, \Delta_-^{-\frac{1}{2}}}{\Delta_{+R}\, \Delta_{-R}^{-\frac{1}{2}}} - \frac{\Delta_-^{\frac{1}{2}}}{\Delta_{-R}^{\frac{1}{2}}} \right) \, w_i(r)\, dx^i\, dv +2\, z_i(r)\, dr \, dx^i\,.
\label{Rvecpara}
\end{equation}	

In the vector sector, we have a constraint equation coming from the spatial components of the energy momentum conservation $\nabla_a T^a_i =0$. Since we are working locally in the neighbourhood of an equilibrium configuration, we obtain the local form of the conservation in terms of the thermodynamic variables $r_+$ and $\beta_i$ as
\begin{align}
-r_+^{13}\; R^8\, \Delta_{+R}\, \Delta_{-R}\, \Delta_-(r_+) \; \partial_v\,\beta_i 
+ \partial_i r_+ \, {\mathfrak v}= 0\,,
\label{vecRcons}
\end{align}	
with
\begin{align}
{\mathfrak v} &= \ell_Q^{16}\, (4\, R^4 - 6\, r_+^4) + \ell_Q^8 \, r_+^4 \, \left(4\, r_+^8- 3\, R^8 + 3\, r_+^4\, R^4\right)+ r_+^{12} \, R^4 \, 
(R^4 - 3\,r_+^4) 
\nonumber \\
& = - \frac{\ell_Q^{20}}{(1- \bp )^2(1-\bE)^{5/2}} \, \left[
6 \, \bp ^2 \, \bE^2 - 8\, \bp ^2 \, \bE - 8 \, \bp  \, \bE^2 +2 \, \bp ^2 +5 \, \bp  \,\bE +2 \, \bE^2 \right] .
\label{veccons1}
\end{align}	
We have expressed the result simply in terms of the physical parameters $\bp $ and $\bE$ introduced in \eqref{pEpars}. Note that the result is symmetric under the exchange of the two parameters.  We will return to the physics of this equation when we discuss the stress tensor in \sec{s:stenfin}.

In addition we have to work out the dynamical equations which will serve to determine the corrections $w_i(r)$ and $z_i(r)$. It transpires that  $z_i(r)$ turns out to be pure gauge -- we set it to zero henceforth. We have only one dynamical equation, which takes the form:
\begin{equation}
\frac{d}{dr}\left(\frac{(R^4 - r^4)^2}{r^3}\, \frac{d w_i}{dr}\right) ={\mathfrak V}_{R} \;\partial_v\beta_i
\label{Rveceq}
\end{equation}	
where ${\mathfrak V}_{R}$ is a total derivative allowing the first integral to be done explicitly. In particular, we find an unilluminating expression for this vectorial source term
\begin{align}
{\mathfrak V}_{R} &= -\frac{1}{{\mathfrak v}} \;
\frac{ R^{12} \, \Delta_{+R}^{\frac{3}{2}}\, \Delta_{-R}^{\frac{7}{4}}}{ r_+^4\, \Delta_{-}(r_+)}  
 \,\frac{d}{dr} \left[\frac{1}{r^3\, \Delta_-^{\frac{5}{4}}}  
\bigg\{
R^4\left(2\,\ell_Q^{16} - r^4\, r_+^4 \left(5\,r_+^8 + \ell_Q^8\right)+ 4\, r_+^8\,\ell_Q^8\right) 
\right.
\nonumber \\
& \left.\qquad \qquad \;\; 
- 8\, r_+^4\, \ell_Q^{16} +r^8 \,r_+^4 \left(r_+^8 -3\,\ell_Q^8\right) + 
r^4 \,\ell_Q^8 \left(2\,\ell_Q^8 + 8\,r_+^8\right)
 \right.
\nonumber \\
& \left.
\;\;\qquad \qquad+ \;
8\, e^{i\frac{\pi}{4}}\, \ell_Q^6\,r^5\, r_+^5\, R^4\, \Delta_{+R}\, \Delta_-^{\frac{5}{4}} \, \;\; _2F_1\left(\frac{1}{4}, \frac{1}{2}, \frac{3}{2}, \frac{r_+^4\, r^4}{\ell_Q^8} \right) 
\bigg\}
\right]\,.
\label{Rvecint1}
\end{align}
Note that the phase conventions are to ensure that we have the appropriate basis of  hypergeometric functions.

Having determined the bulk equations, we turn to the boundary conditions. Our parameterization of the metric correction 
 in \eqref{Rvecpara} has naturally ensured that the perturbation vanishes on the Dirichlet surface $r = R$. One can check that the source term ${\mathfrak V}_R$ is regular on the horizon (it must be since the background geometry is smooth there). So we a-priori have 2 independent parameters which we can take w.l.o.g. to be $w_i(R)$ and $w_i'(R)$. 
Of these two parameters,  one is fixed by regularity at the Dirichlet surface: since the differential operator in \eqref{Rveceq} has a double zero at $r=R$ we need to fix $w'_i(R)$ appropriately. It turns to that the choice given by the following first integral of the equation
\begin{equation}
\frac{dw_i(r)}{dr} =\partial_v\beta_i \; \frac{r^3}{(R^4 - r^4)^2} \;\int_r^{R}\; d\rho\; {\mathfrak V}_{R}(\rho)
\label{}
\end{equation}	
serves to ensure that the function remains well behaved as $r \to R$.

To summarize, in the vector sector we have an explicit solution parameterized by constants $w_i(R)$ which serve to characterize a gauge choice.\footnote{
The astute reader might wonder if a similar issue does not arise from the parameter $z_i(R)$ which could be non-zero since $z_i(r)$ is pure gauge. This term does not contribute to the stress tensor at the order we are working in and hence can be ignored.}
 As we explain in \sec{s:stenfin}, the stress tensor obtained from this perturbation whilst generically of the hydrodynamic form, is in Landau frame only for a specific choice of this parameter. We therefore leave it arbitrary for the moment and will fix $w_i(R)$ when we compute the world-volume stress tensor. In any event, since it serves to determine only the choice of fluid frame, we can leave it arbitrary for all physical purposes (and concentrate on frame-invariant physical data).   Finally, we define again, for future convenience, a function $\mathbf{v}$ for the vector perturbation solution 
\begin{equation}
{\mathbf v}(r,\sigma^a) \, \partial_v\beta_i \equiv \left(\frac{\Delta_+\, \Delta_-^{-\frac{1}{2}}}{\Delta_{+R}\, \Delta_{-R}^{-\frac{1}{2}}} - \frac{\Delta_-^{\frac{1}{2}}}{\Delta_{-R}^{\frac{1}{2}}} \right) \, w_i(r)\,.
\label{vfn}
\end{equation}	
%

\subsubsection{Scalars of $SO(3)$}
\label{s:finscalar}

Finally we move on to the scalar sector where a parameterization of the metric correction can be taken to be
\begin{equation}
ds^2_{(1),S} =   k(r)\, dv^2 +2\, \frac{ \Delta_-^{-\frac{3}{4}}}{\Delta_{-R}^{-\frac{3}{4}}\; \Delta_{+R}^{\frac{1}{2}}} \, j(r) \, dv \,dr +
 \frac{\Delta_-^{\frac{1}{2}}}{\Delta_{-R}^{\frac{1}{2}}} \;h(r)\, dx^i dx^i\,.
\label{asvecpar}
\end{equation}	
Note that since we have used the dilaton to gauge fix our coordinates, we should allow for three functions in the metric correction (this is the correct degrees of freedom counting). This turns out to be the most complex sector to analyze. A-priori we expect three dynamical equations and one conservation equation in this sector.

 The conservation equation is obtained by looking at the scalar equation $\nabla_a T^a_v =0$ which in local fluid variables comes from the linear combination involving $\text{Eom}_{vr}$ and $\text{Eom}_{vv}$ components of Einstein's equations. This reads:
\begin{equation}
R^4\,r_+^{13} \, \Delta_{-R}\, \Delta_-(r_+) \, \partial_i\,\beta_i  
+ \partial_v r_+ \; {\mathfrak s}
= 0\,,
\label{scaRcons}
\end{equation}	
where we define
\begin{align}
{\mathfrak s} &= R^4\,r_+^4 \left(5\,r_+^8 + \ell_Q^8\right) + 2\, \ell_Q^8 \left(\ell_Q^8 - 4\, r_+^8\right)\,,
\nonumber \\
& =\frac{\ell_Q^{16}}{(1-\bp)(1-\bE)^2 } \, \left[
 3 \, \bE+6 \,\bp +2 \, \bE^2-4 \, \bE \, \bp-2 \, \bE^2 \, \bp   
\right] 
\label{fraks}
\end{align}	
for future notational simplicity.

 To get the dynamical equations, one needs to play around with suitable linear combinations of the equations of motion. A-priori an obvious basis of equations to chose in the naked frame\footnote{
 The basis of equations in Einstein frame is slightly modified, but the upshot is the same as what we derive in the naked frame described here.} involves $\text{Eom}_{rr}$, $\text{Eom}_\varphi$ and $\text{Eom}_\mu^\mu$ (the trace of Einstein's equation). However, we find it useful to take for the last equation a linear combination of $\text{Eom}_\mu^\mu$ and $\text{Eom}_{vr}$ to simplify the analysis.
 
 The first two equations are pretty simple to obtain. We have
\begin{equation}
\text{Eom}_{rr} =0 : \qquad j' = \frac{3}{2\,\Delta_{-R}^{\frac{1}{2}}} \; \frac{r^5\, r_+^4\;\Delta_-^{-\frac{1}{4}}}{5\,r^4\,r_+^4 -2\, \ell_Q^8} \left(\Delta_-^{\frac{5}{4}} \, h'\right)'
\label{Rrreq}
\end{equation}	
and
\begin{align}
\text{Eom}_\varphi &=0: \quad 
 \left(\Delta_-^{\frac{3}{2}} \, r^4 \;k\right)' &= 
\frac{1}{2}\; r^4\, \Delta_+\,\Delta_- \; \frac{ \Delta_{-R}^{\frac{1}{2}}}{\Delta_{+R}} 
\left(3\, h' -2 \, \Delta_{-R}^{\frac{1}{2}}\, j'\right) - 8\, \frac{r^8-\ell_Q^8}{r^5} \, \frac{\Delta_{-R}}{\Delta_{+R}} \; j + {\mathfrak S}_1
\label{Rphieq}
 \end{align}
where ${\mathfrak S}_1$ is the source term given by
\begin{align}
{\mathfrak S}_1 & = \partial_i\, \beta_i  \; \frac{R^4}{{\mathfrak s}}  \left(\frac{\Delta_{-R}^{\frac{5}{4}}}{\Delta_{+R}^{\frac{1}{2}}}\right)
 \left[\frac{ 2\,\ell_Q^{16}+ 5\, r^4\, r_+^{12} + \ell_Q^8\,r_+^4 \left(r^4-8\,r_+^4\right)}{\Delta_-^{\frac{1}{4}}}  \right]
 \nonumber \\
 & = \partial_i\, \beta_i  \; \frac{R^4\, r_+^4}{{\mathfrak s}}  \left(\frac{\Delta_{-R}^{\frac{5}{4}}}{\Delta_{+R}^{\frac{1}{2}}}\right)
 \left[\frac{r_+^4\, \Delta_-(r_+) \, \left(5\, r^4\, r_+^4 - 2\, \ell_Q^8\right) + 6\, \ell_Q^8\,  r^4\, \Delta_+
 }{\Delta_-^{\frac{1}{4}}}  \right] ,
\label{}
\end{align}
where we have written the second line in terms of functions that naturally appear in the background metric.

The third equation we use is obtained as mentioned from $\text{Eom}^\mu_\mu$ and $\text{Eom}_{vr}$. This reads:
\begin{align}
\left(\Delta_-^{\frac{1}{2}} \;k\right)' &=  {\mathfrak S}_2
- \left( \frac{\Delta_{-R}}{\Delta_{+R}}\right) \left(
\frac{8\, r_+^4}{r^5}    \; j(r) 
+ \frac{\Delta_+ \, (\ell_Q^8 + 5\, r^4\, r_+^4)}{3\,\ell_Q^8} \; j'(r) \right) 
\nonumber \\
& 
+ \frac{\Delta_{-R}^{\frac{1}{2}}}{\ell_Q^8\, \Delta_{+R}} \left(r_+^4\, r^5\  \Delta_+\, \Delta_- \; h''(r)+ \frac{5\, r^8\, r_+^4 - r^4\, r_+^8+ \ell_Q^8 ( 7\, r^4 - 11\,r_+^4)}{2\,r^4} \,  h'(r) \right)
\label{R2ndeq}
\end{align}
with a source
\begin{align}
{\mathfrak S}_2 &= \frac{\partial_i\beta_i}{3\,\ell_Q^8\; {\mathfrak s}} \; \frac{R^4}{r^4} \left( \frac{\Delta_{-R}^{\frac{5}{4}}}{\Delta_{+R}^{\frac{1}{2}}} \right) \; \frac{2\,\ell_Q^{24} - \ell_Q^{16}\,r_+^4\left(13\,r^4+ 8\,r_+^4\right) + \ell_Q^8\,r_+^8\, r^4\left(5\, r^4 - 11\, r_+^4\right)+ 25\, r_+^{16}\, r^8}{\Delta_-^{\frac{5}{4}}}
\nonumber \\
& =\left(\frac{\ell_Q^8+5\,r^4\,r_+^4}{3\,\ell_Q^8\,r^4 \, \Delta_-}\right) {\mathfrak S}_1 + \frac{\partial_i\beta_i}{ {\mathfrak s}} \left( \frac{\Delta_{-R}^{\frac{5}{4}}}{\Delta_{+R}^{\frac{1}{2}}} \right) \; \frac{8\,  r_+^{12}\, r^4\, \Delta_-(r_+)
}{\Delta_-^{\frac{5}{4}}}\,,
\label{}
\end{align}	
which has been written so as to make a contribution proportional to ${\mathfrak S}_1$ manifest.

To simplify this system we consider the system of equations \eqref{Rphieq} and \eqref{R2ndeq} as determining $k(r)$ algebraically once we eliminate from them $k'(r)$, which is easily done. Since $k(r)$ is algebraically determined in terms of $\{j(r),h'(r),h''(r)\}$ we can use it back in \eqref{Rphieq} to obtain an equation involving the functions $j(r)$ and $h(r)$ alone. A-priori we expect to see the appearance of $\{j,j',h',h'',h'''\}$ in the resulting equation. However, this equation actually ends up providing us with our first simplification: $j(r)$ cancels out from the expression entirely. Since \eqref{Rrreq} allows us to determine $j(r)$ from $\{h',h''\}$ we can finally obtain an autonomous equation for $h(r)$.

We end up with  an inhomogeneous second order ODE for $h'(r)$ (noting that $h(r)$ never appears in the equations of motion undifferentiated), which we can deal with analytically. To present this compactly, let:
\begin{equation}
h'(r) \equiv \frac{1}{r^5\, \Delta_-^2}\, q(r) 
\label{qdefn}
\end{equation}	
which results in
\begin{equation}
\frac{d}{dr}\left(\frac{r^9\, \Delta_-\, \Delta_+^2}{5\,r^4\,r_+^4 - 2\,\ell_Q^8} \; q'(r) \right)
= {\mathfrak S}_q
\label{Rpeq}
\end{equation}	
with a new source term
\begin{equation}
{\mathfrak S}_q =- {\mathfrak s}_q\; \frac{\Delta_+}{\Delta_-^{\frac{1}{4}}}  
\bigg[4\, \ell_Q^{24} -16\, \ell_Q^{16}\,r_+^4\, \left(r^4 + r_+^4\right)+ \ell_Q^8 \, r^4\, r_+^8\, \left(9\, r^4 - 26\, r_+^4\right) + 45\, r^8\,r_+^{16} \bigg]
\label{}
\end{equation}	
and for convenience we introduce
\begin{equation}
{\mathfrak s}_q = \frac{2\,R^4\, \Delta_{+R}^{\frac{1}{2}}\, \Delta_{-R}^{\frac{3}{4}}}{ 3\,r_+^{8}\, {\mathfrak s}}\;\partial_i\beta_i \,.
\label{}
\end{equation}	

Now the good news is that ${\mathfrak S}_q$ is integrable; and the associated constant of integration is also fixed. This follows from the from the kinetic term in \eqref{Rpeq} which has a factor $\Delta_+^2$ -- we therefore require that $\int dr\, {\mathfrak S}_q \to (r-r+)^2$ near the horizon. This determines the first of our integration constants which we denote as $C_{q_{1}}$.  Then by using \eqref{qdefn} we learn that $h(r)$ is regular on the horizon, which ensures regularity of the other two functions $k(r)$ and $j(r)$. With the regularity implemented (we don't give the expression for $C_{q_1}$) we find then that 
\begin{align}
q'(r) &= {\mathfrak s}_q \, r_+^4\, \frac{5\,r^4\,r_+^4 - 2\,\ell_Q^8}{r^9\, \Delta_-\, \Delta_+^2}
 \left[- 4 \,r_+^{21}\, \, \Delta_{-}(r_+)^{\frac{11}{4}}  \right.
 \nonumber \\
 & \left. \qquad+\; r\, \Delta_-^{\frac{3}{4}} \, 
\left( -5\, r^8\,r_+^{12} + 9\, r^4\, r_+^{16}  -\ell_Q^8 \,r_+^4\left(
r^8-r^4\,r_+^4 + 8\,r_+^8\right)+ 2\,\ell_Q^{16}\, (r^4+r_+^4) \right)\right] .
\label{}
\end{align}	

To proceed further, we need the following integrals in some useful form:
\begin{align}
{\cal I}_1  &= \int dr\; \frac{5\,r^4\,r_+^4 - 2\,\ell_Q^8}{r^9\, \Delta_-\, \Delta_+^2}
\nonumber \\
{\cal I}_{2(n)}  &= \int dr\; \frac{5\,r^4\,r_+^4 - 2\,\ell_Q^8}{r^{4n}\, \Delta_-^{\frac{1}{4}}\, \Delta_+^2} =  \int dr\; \frac{r_+^4}{r^{4n}\, \Delta_+^2}\,\left(r^5\,\Delta_-^{\frac{3}{4}}\right)'
\label{}
\end{align}	
since we can then write 
\begin{align}
q(r) &=C_{q_{2}}+ r_+^4\, {\mathfrak s}_q 
\left[- 4 \,r_+^{21}\, \, \Delta_{-}(r_+)^{\frac{11}{4}} \, {\cal I}_1 + 2\,\ell_Q^8\, r_+^4\, (\ell_Q^8-4\,r_+^8)\, {\cal I}_{2(2)} \right.
\nonumber \\
& \left. \qquad 
+ \;(2\,\ell_Q^{16}+ \ell_Q^8\,r_+^8 +9\,r_+^{16}) \, {\cal I}_{2(1)}- r_+^4\, (\ell_Q^8+ 5\,r_+^8)\, {\cal I}_{2(0)} \right]\nonumber \\
& \equiv C_{q_{2}}+ r_+^4\, {\mathfrak s}_q \left[- 4 \,r_+^{21}\, \, \Delta_{-}(r_+)^{\frac{11}{4}} \, {\cal I}_1 + {\cal I}_2\right]\,.
\label{qfinal}
\end{align}	

It turns out that the expression for the integrals can be obtained in terms of ordinary logarithms and  Appell hypergeometric functions. For instance:
\begin{equation}
{\cal I}_1 = \frac{1}{4\, r_+^8\, \Delta_-(r_+)^2} \left[ \frac{-2\,\ell_Q^{16} + 7\, \ell_Q^8\,r_+^8 - 5\, r_+^{16}}{r_+^4} \; \frac{1}{r^4\, \Delta_+} + 3\, \ell_Q^8 \, \log \left(\frac{\Delta_-}{\Delta_+} \right) \right] \,.
\label{}
\end{equation}	
While each of the individual expression for ${\cal I}_{2(n)}$ can be obtained in a reasonable form, the particular linear combination that appears in the expression for $q(r)$ in \eqref{qfinal}, which we defined to be ${\cal I}_2$ is actually simpler.
\begin{align}
{\cal I}_2 &= - r_+^4 \, r\, \left(2\,\ell_Q^{16} + \ell_Q^8 \,r_+^4 \,(r^4 - 8\, r_+^4) + 5\, r_+^{12}\, r^4\right) \, \frac{\Delta_-^{\frac{3}{4}}}{\Delta_+} 
\nonumber \\
& \qquad +\; 2\, \ell_Q^6\, r_+^{9}\, \Delta_-(r_+) \, r^6\, e^{i\,\frac{\pi}{4}}\, \text{AppellF}_1\left(\frac{3}{2}, \frac{1}{4},1,\frac{5}{2},\frac{r_+^4\, r^4}{\ell_Q^8} , \frac{r^4}{r_+^4} \right)\,.
\label{}
\end{align}	
The phase conventions here ensure that we are on the physical branch of the solutions.

Putting these together, we learn that 
\begin{align}
q(r) &= C_{q_2} + r_+^4\, {\mathfrak s}_q \left[-r_+^{13}\, \Delta_-(r_+)^{\frac{3}{4}}\, 
\left[ \frac{-2\,\ell_Q^{16} + 7\, \ell_Q^8\,r_+^8 - 5\, r_+^{16}}{r_+^4} \; \frac{1}{r^4\, \Delta_+} + 3\, \ell_Q^8 \, \log \left(\frac{ \Delta_-}{\Delta_-} \right) \right] \right.
\nonumber \\
&\left. 
\qquad -\;r_+^4 \, r\, \left(2\,\ell_Q^{16} + \ell_Q^8 \,r_+^4 \,(r^4 - 8\, r_+^4) + 5\, r_+^{12}\, r^4\right) \, \frac{\Delta_-^{\frac{3}{4}}}{\Delta_+} 
\right.
\nonumber \\
& \left.
\qquad +\; 2\, \ell_Q^6\, r_+^9\, \Delta_-(r_+) \, r^6\, e^{i\,\frac{\pi}{4}}\, \text{AppellF}_1\left(\frac{3}{2}, \frac{1}{4},1,\frac{5}{2},\frac{r_+^4\, r^4}{\ell_Q^8} , \frac{r^4}{r_+^4} \right) \right]\,.
\label{}
\end{align}

Having determined $q(r)$ we therefore are left with having to solve \eqref{qdefn} for $h(r)$ from which we can then determine $k(r)$ and $j(r)$. As such at this point we still have $C_{q_2}$ which is unfixed and we expect another constant of integration $C_{q_3}$ when we integrate \eqref{qfinal} for $h(r)$. Since the Dirichlet boundary condition imposes on us $h(R) =0$, it is clear that $C_{q_3}$ can be fixed in terms of  $C_{q_2}$, so we only have a one parameter ambiguity in the problem. To solve for $j(r)$, we integrate \eqref{Rrreq} which results in yet another integration constant; w.l.o.g. we can refer to this as $j(R)$. As in the case of the vector modes discussed in \sec{s:finvector} we find that the Dirichlet boundary condition  at the surface $r=R$ is insensitive to the value of $j(R)$, which forms another free parameter for our problem.

Finally, $k(r)$ is, as we have argued, determined algebraically in terms of the functions $\{j(r),j'(r), h'(r),h''(r)\}$ (by differencing \eqref{Rphieq} and \eqref{R2ndeq}), so there is no new integration constant to be obtained. However, it is clear that $k(R)$ depends on $C_{q_2}$ and $j(R)$ along with potential source terms coming from the explicit evaluation of the other functions. Since we demand $k(R) = 0$, we can use this condition to fix $C_{q_2}$. The upshot is that we have, after imposition of the Dirichlet boundary condition, a one-parameter family of bulk solutions satisfying the equations of motion, which we are going to label using $j(R)$. The computation of the hypersurface stress tensor will provide us with a condition to fix $j(R)$ uniquely as we will see in \sec{s:stenfin}. To wrap things up, let us finally introduce three functions $\mathbf{s}_i$ obtained from the scalar perturbation solution,
\begin{equation}
{\mathbf s}_1(r,\sigma^a)\, \partial_i\beta^i \equiv k(r)\,,
\qquad 
{\mathbf s}_2(r,\sigma^a) \,\partial_i\beta^i \equiv  3\, \frac{\Delta_-^{\frac{1}{2}}}{\Delta_{-R}^{\frac{1}{2}}} \;h(r)
 \qquad 
{\mathbf s}_3(r,\sigma^a) = \frac{ \Delta_-^{-\frac{3}{4}}}{\Delta_{-R}^{-\frac{3}{4}}\; \Delta_{+R}^{\frac{1}{2}}} \, j(r) \,.
\label{sifn}
\end{equation}	
%

\subsubsection{Summary of perturbation analysis}
\label{s:sumg1}

The perturbation analysis carried out for non-linear long-wavelength fluctuations of the D3-brane geometry 
results in a solution to the supergravity equations of motion valid to first order in gradient expansions. We solved the dynamical equations of motion ($\text{Eom}_{ab}$, $\text{Eom}_{ar}$ and $\text{Eom}_\varphi$) and determined the constraints that the fields $u^a$ (equivalently $\beta_i$) and $r_+$ are required to satisfy, 
\eqref{vecRcons}  and \eqref{scaRcons}. The bulk metric to this order is of the form
\begin{equation}
g_{\mu\nu}\,dx^\mu\, dx^\nu = g^{(0)}_{\mu\nu}\,dx^\mu\, dx^\nu +  \epsilon \, g^{(1)}_{\mu\nu}\,dx^\mu\, dx^\nu
\label{}
\end{equation}	
with $g^{(0)}_{\mu\nu}$ having been presented in a world-volume covariant form in \eqref{5dseed}. The first order correction can be written in a covariant form:
\begin{equation}
g^{(1)}_{\mu\nu}\, dx^\mu \, dx^\nu = {\mathbf t} \, \sigma_{ab}\, dx^a \, dx^b +   \left[\theta\, \left(
{\mathbf s}_1 \, u_{a}\, u_{b}\, + \frac{1}{3}\, {\mathbf s}_2\, P_{ab}\right)  + {\mathbf v} \, a_{(a}\, u_{b)} \right] dx^a \, dx^b + 2\, {\mathbf s}_3\, \theta \, u_a\, dx^a\, dr
\label{}
\end{equation}	
with
\begin{equation}
\sigma_{ab} = P_a^c\, P_b^d\, \left(\nabla_{(c} u_{d)} - \frac{1}{3}\, P_{cd}\, \theta\right) \,,\qquad \theta = \nabla_a u^a\,,\qquad a^c = u^b\,\nabla_b u^c 
\label{}
\end{equation}	
being the shear tensor,  expansion, and acceleration of the fluid respectively. The functions ${\mathbf t},\; {\mathbf s}_i,\; {\mathbf v} $ can be read off from \eqref{tfn}, \eqref{vfn}, \eqref{sifn} (we refrain from writing out the explicit forms since the final expressions are unilluminating). All we have done is isolate the explicit dependence on the gradients of the fluid velocity. The $r_+$ gradient is fixed in terms of the gradients of $u^a$ by the constraint equations which we will examine in some detail shortly.

We note that our solution derived so far is parameterized by two sets of parameters, $w_i(R)$ in the ${\bf 3}$ of $SO(3)$ which was left arbitrary in the vector sector, and $j(R)$ which was left undetermined in the scalar sector.

\subsection{The boundary stress tensor}
\label{s:stenfin}

We now have all the ingredients necessary to compute the stress tensor induced on the brane world-volume $r = R$. 
To compute this stress tensor we use the five-dimensional effective action with all the boundary terms and counter-terms described in \sec{s:5deff}. While it is a straightforward exercise to compute the extrinsic curvature of the 
Dirichlet hypersurface  and the other intrinsic data, it is useful to write down a somewhat abstract result for the stress tensor first. This has the merit of allowing us to explain how to fix the remaining parameters characterizing our solution, viz., $w_i(R)$ and $j(R)$. 

Let us start by writing the stress tensor given in \eqref{Tab2} in the gradient expansion as:
\begin{equation}
 T_{ab} = \frac{1}{\kappa_5^2}\left(\, R^5 \left(\Theta_{ab}-h_{ab}\Theta \right) +\left[ 5\, R^4\, \left(e^\varphi \, \sqrt{(\partial_n \varphi)^2} -1 \right) + Q \right] h_{ab} \right) \equiv T_{ab}^{(0)} + T_{ab}^{(1)} + \cdots
\label{Tab01}
\end{equation}	
where $ T^{(k)}_{ab}$ is the $k^{\rm th}$ order result for the stress tensor. As should be clear $T^{(0)}_{ab}$ can in fact be quite easily evaluated already from \eqref{5dsoln}. One has:
\begin{equation}
T_{ab}^{(0)}  = \varepsilon_Q\, u_a\, u_b + P_Q\,P_{ab}   
\label{}
\end{equation}	
with local energy density and pressure given by
\begin{align}
 \varepsilon_Q &=  \frac{R^4}{\kappa_5^2}\left(5-\frac{Q}{R^4} 
 -5\, \sqrt{\Delta_{-R}\, \Delta_{+R}} - \frac{3}{4}\,R\; \sqrt{\frac{\Delta_{+R}}{\Delta_{-R}} }\, \Delta_{-R}' \right)\,, \nonumber\\
 P_Q &=-   \varepsilon_Q + \frac{R^5} 
 {\kappa_5^2}\; \frac{\Delta_{-R}\, \Delta_{+R}' - \Delta_{+R}\, \Delta_{-R}'}{2\,\sqrt{\Delta_{+R}\, \Delta_{-R}}} \,,
\label{zeroSTv1}
\end{align}
where all the functions and their derivatives are evaluated at $r =R$. Note that here $r_+$ and $u_a$ are now functions of $\sigma^a$, so the part of the stress tensor computed above does in fact give a piece at first  order in the gradients. However, we know their functional form is constrained to be obtained via a Taylor expansion in the gradients of these fields.  Given these expressions for the energy density and the pressure, it is a simple exercise to write down the leading order equation for the conservation of the stress tensor $\nabla^a T^{(0)}_{ab} = 0$. The resulting equations are nothing but the gravitational constraint equations \eqref{vecRcons} and \eqref{scaRcons} derived in \sec{s:d3pert}. 
In terms of our simplified variables $\bE$ and $\bp$ in \eqref{pEpars}, the energy density and pressure are given by
\begin{align}
 \varepsilon_Q &=  \frac{\ell_Q^4}{\kappa_5^2}\frac{1}{\sqrt{1-\bE}}\left(
 \frac{5}{1-\bp}
 -2\, \sqrt{1-\bE} 
-5 \, \frac{\sqrt{\bp \, (\bE + \bp - \bE \, \bp)}}{1-\bp}
- 3 \, (1-\bE) \, \sqrt{\frac{\bp}{\bE + \bp - \bE \, \bp}}
  \right) \nonumber \\
 P_Q &=-\varepsilon_Q+
 \frac{\ell_Q^4}{\kappa_5^2}
\frac{ 2 \, \bE }{\sqrt{\bp \, (\bE + \bp - \bE \, \bp)(1-\bE)}}
  \ .
\label{zeroST}
\end{align}
%
\begin{figure}[h!]
\begin{center}
\includegraphics[width=6in]{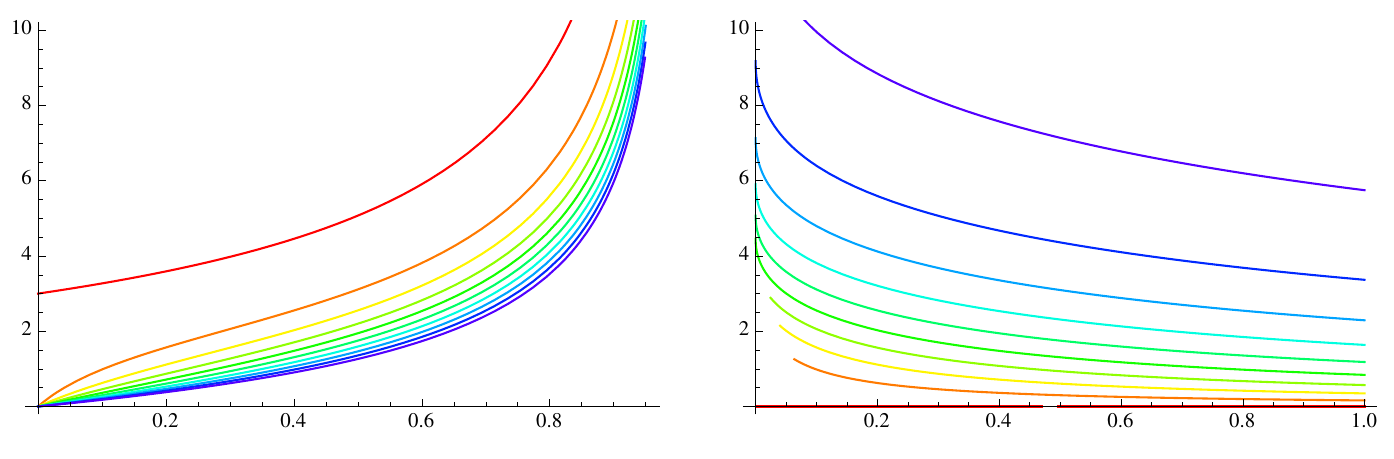}\\
\vspace{1cm}
\includegraphics[width=6in]{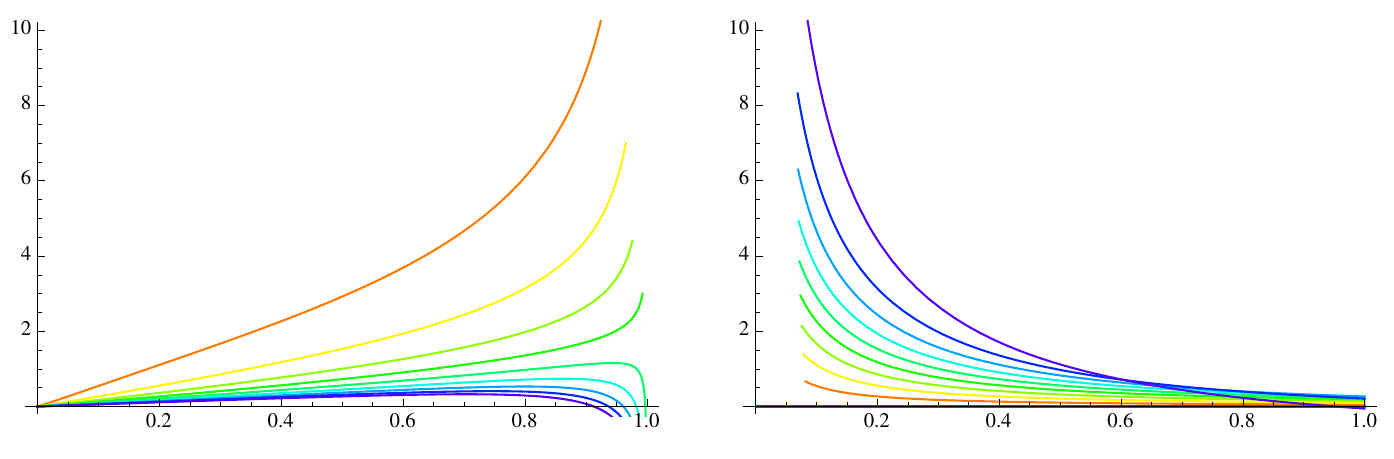}
\begin{picture}(0,0)
\setlength{\unitlength}{1cm}
\put (-0.1,0.5) {$\bp$}
\put (-7.7,5) {$P_Q$}
\put (-8,0.5) {$\bE$}
\put (-15.7,5) {$P_Q$}
\put (-0.2,6) {$\bp$}
\put (-7.7,11) {$\varepsilon_Q$}
\put (-8,6) {$\bE$}
\put (-15.8,11) {$\varepsilon_Q$}
\put(-12,8.5){$\downarrow \,\bp$}
\put(-12,2){$\downarrow \,\bp$}
\put(-6.8,7.5){$\uparrow \,\bE$}
\put(-7,2){$\uparrow \,\bE$}
\end{picture}
\caption{
Plots of the energy density $\varepsilon_Q$ and pressure $P_Q$ (in units of  $\ell_Q^4/\kappa_5^2$).
On the left panels we show the behaviour as a function of the extremality parameter $\bE$ with the individual curves corresponding to different values of the cutoff surface placement $\bp$.  The color coding (online) is such that $\bp \to 0$ (IR or horizon) is red and $\bp \to 1$ (UV or asymptopia) is purple. On the right panels the behaviour is illustrated as a function of $\bp$. Now $\bE=0$ corresponding to the extremal brane is red while the $\bE =1$ corresponding to the neutral brane is purple.  While the information contained in the left and right panels is redundant we present it for ease of visualization. Note that the finite energy density at extremality on the horizon $\bp =0$ is merely due to an order of limits.
} 
\label{f:epsQPQ}
\end{center}
\end{figure}
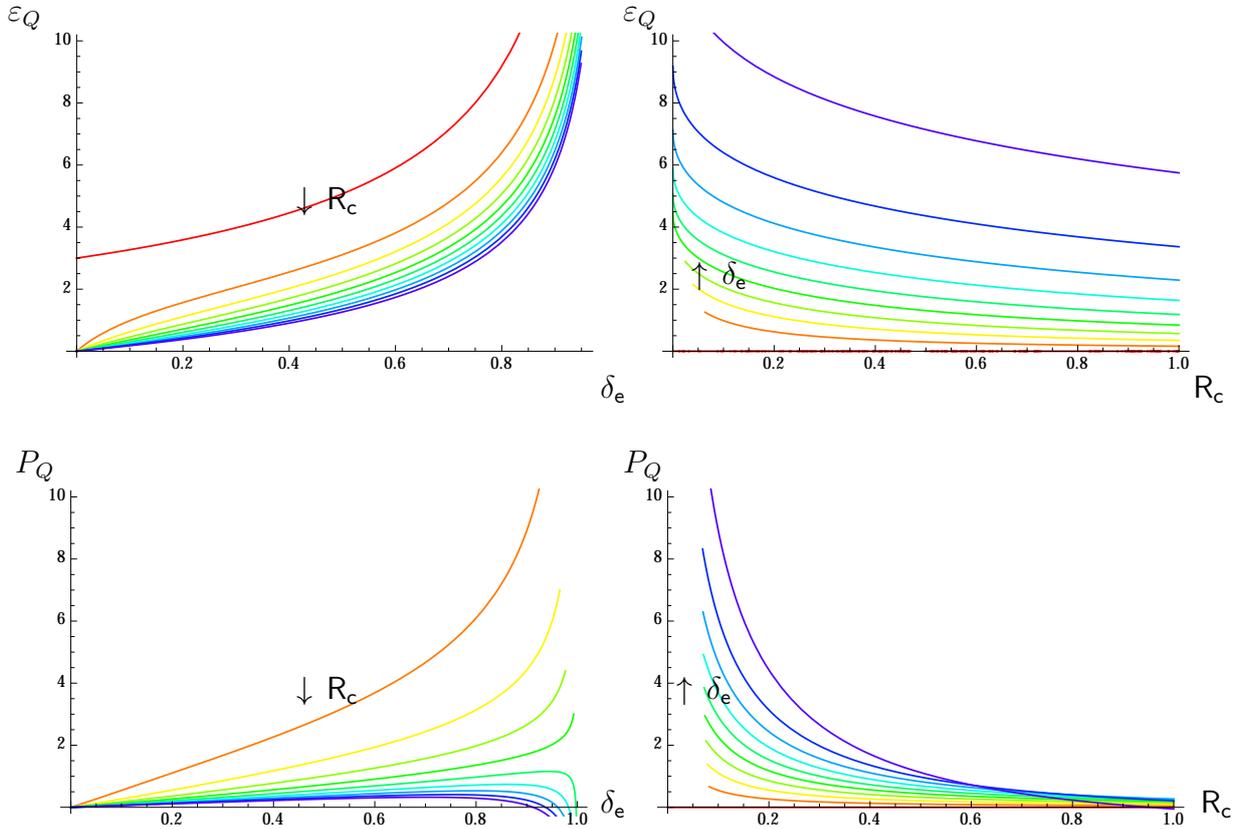
For illustration, in \fig{f:epsQPQ} we show the behaviour of $\varepsilon_Q$ and  $P_Q$ as functions of  $\bE$ and $\bp$.
We can easily check that the energy density is positive everywhere in the entire range of parameters, $0<\bp<1$ and $0<\bE<1$.  On the other hand, the  pressure  becomes negative for $\bE$ and $\bp$ sufficiently close to 1.  This might lead us to expect that near-neutral branes in asymptotically large box are unstable -- which is indeed the case as we'll see below.

The second term on the  r.h.s.\  of \req{Tab01},
$T^{(1)}$, on the other hand is genuinely of first order in the gradient expansion and corresponds to the viscous stress tensor. Since we are holding the net D3-brane charge fixed in the physical processes we are considering, we should expect that $T^{(1)}_{ab}$ should take the form of a viscous stress tensor for a neutral fluid, i.e.,
\begin{equation}
T^{(1)}_{ab} = -2\, \eta_Q\, \sigma_{ab} - \zeta_Q\, \theta\, P_{ab}
\label{}
\end{equation}	
where $\sigma_{ab}$ is the shear and $\theta$ the expansion of the fluid velocity $u_a$. In writing the above we have assumed that we are working in the Landau frame for the fluid dynamics, i.e., $u^a$ is a (normalized) timelike eigenvector of $T_{ab}$. This implies that $T^{(k)}_{ab} \, u^a = 0$ for $k>0$. 
Should we not be working with such a frame choice for the fluid, then at the first order in gradients we would generally expect to see two other terms:
\begin{equation}
\left(T^{(1)}_{ab}\right)_{\text{non-Landau}} = -2\, \eta_Q\, \sigma_{ab} - \zeta_Q\, \theta\, P_{ab} - \kappa_Q\, a_{(a} \, u_{b)} - \zeta'_Q\, u_a\, u_b \,\theta
\label{nlst}
\end{equation}	
where $\kappa_Q$ is the heat conductivity and $\zeta'_Q$ is a shift of the local energy density by the expansion of the fluid. We can demand that the hypersurface stress tensor at $r=R$ is in the Landau frame, by an appropriate choice of $u^a$ (frame choices in fluid dynamics are of course obtained by field redefinitions). From the point of view of the bulk solution obtained in the previous sections, this frame choice can be implemented by choosing our integration constants appropriately. 

Note that we had an  ambiguity in our solution for $g^{(1)}$: the values of the functions $w_i(R)$ in the vector of $SO(3)$ and $j(R)$ in the scalar of $SO(3)$ at $r =R$ was the freedom we isolated. The ambiguity in \eqref{nlst} is also precisely of this form (as it must be on symmetry grounds): $\kappa_Q$ is a vector and can be eliminated by an appropriate choice of $w_i(R)$ while $\zeta'_Q$ is a scalar and can be set to zero by a choice of $j(R)$. We note that this was guaranteed: one can work with frame-invariant fluid dynamical data and realize that a neutral fluid has precisely one frame-independent tensor and one frame-independent scalar data, which can w.l.o.g. be identified with $\eta_Q$ and $\zeta_Q$ respectively, cf., \cite{Bhattacharya:2011eea} for a discussion.

By an explicit computation of the stress tensor on the solution we find a relation that needs to be satisfied between $j(R)$ and $h'(R)$
\begin{equation}
j(R) = \frac{2\, R}{\sqrt{\Delta_{+R}}} \; \frac{2\,\theta + 3\, \sqrt{\Delta_{+R}\,\Delta_{-R}} \; h'(R)}{20\, \Delta_{-R} + 3\, R\, \Delta_{-R}'} \ .
\label{}
\end{equation}	
This relation suffices to set $\zeta'_Q$ to zero. However, since there is non-trivial information in the scalar sector given by the bulk viscosity, the precise value of $j(R)$ also affects $\zeta_Q$. A similar exercise picks out a value for $j_i'(R)$; since its value does not enter any other component of the stress tensor (there being no frame invariant vector data for a neutral fluid), we will refrain from quoting its (unedifying) value required to set $\kappa_Q = 0$.

Having fixed the parameters we can read off by an explicit computation the physically meaningful values of the fluid dynamical transport coefficients. We find: 
\begin{equation}
\eta_Q = \frac{1}{2\, \kappa_5^2}\;  r_+^5\; \frac{\Delta_-(r_+)^{\frac{3}{4}}}{\Delta_{-R}^{\frac{3}{4}}} = \frac{\ell_Q^5}{2\, \kappa_5^2}\; 
 (1-\bE)^{-5/8} \, \left( \frac{\bE}{\bE + \bp - \bE \, \bp} \right)^{\! \frac{3}{4}} \,,
\label{etaD3}
\end{equation}	
\begin{align}
\zeta_Q &= \frac{40}{3\, \kappa_5^2}\;  \frac{R^8\, r_+^{29}}{{\mathfrak s}^2} \, \Delta_{-R}^{\frac{5}{4}} \, \Delta_-(r_+)^{\frac{11}{4}}
= \frac{40}{3\, \kappa_5^2}\;  \frac{\ell_Q^{37}}{{\mathfrak s}^2\, \left(1-\bp\right)^2} \,
\frac{
\bE^{\! \frac{11}{4}}\, \left(\bE + \bp - \bE \, \bp\right)^{\! \frac{5}{4}}}{(1-\bE)^{37/8}} \,,
\label{zetaD3}
\end{align}	
in terms of the variables  $\bp$, $\bE$, ${\mathfrak s}$ defined in \eqref{pEpars} and \eqref{fraks}.

Let us finally note that the speed of sound in this system is obtained from the constraint equations \eqref{vecRcons} and \eqref{scaRcons}. Solving these for plane wave modes of $\beta_i = \delta \beta\, e^{-i\,\omega \, v + i\, k\,x}$ and $r_+ = r_+^{(0)} + \delta r_+\, e^{-i\,\omega \, v + i\, k\,x}$ we obtain the speed of sound:
\begin{align}
v_s^2 &= -\frac{{\mathfrak v}}{{\mathfrak s}}\, \frac{1}{R^4\, \bp}
= \frac{2\,\bE^2 + 5\, \bE\, \bp - 8\, \bE^2\, \bp + 2\,\bp^2 - 8\, \bE\, \bp^2 + 6\, \bE^2\,\bp^2}{\bp \left(3\,\bE + 3\, \bE^2 + 6\,\bp - 4\,\bE\,\bp -2 \, \bE^2\,\bp  \right)}\, .
\label{soundrho1}
\end{align}	
While we have written the final expression in terms of $\bE$ and $\bp$ we should remind the remind the reader that the physical parameters that are held fixed are the location of the cutoff surface, $R$  and the D3-brane charge given by $\ell_Q$.

It is useful to view this expression in terms of the original parameters of the black hole $r_{\pm}$ instead of working with the auxiliary length scale $\ell_Q$ we introduced to simplify the computation. One finds:
\begin{equation}
v_s^2 = \frac{r _-^8 \left(4 \,R^4-6 \,r_+^4\right)+r_-^4 \left(4 \,r_+^8+3 \,r_+^4 \,R^4-3 \,R^8\right)+r_+^4\, R^8-3\, r_+^8\, R^4}{\left(r_+^4-R^4\right) \left(2\,r_-^8+r_-^4 \left(R^4-8\, r_+^4\right)+5 \,r_+^4\, R^4\right)} \,.
\label{soundrho2}
\end{equation}	
%

\section{Salient features of the D3-brane hydrodynamics}
\label{s:physics}

We are now in a position to discuss the physical aspects of our analysis, having dwelt in some detail on the technical aspects of our construction  in the preceding sections. The D3-brane fluid is a viscous relativistic fluid with energy density and pressure given in \eqref{zeroST} and the viscosities are given in \eqref{etaD3} and \eqref{zetaD3}. Note that we are keeping the net charge of the D3-brane from fluctuating and as a result we obtain the dynamics of a neutral fluid. The D3-brane charge does not add any hydrodynamic degree of freedom; it enters in the equation of state and in the constitutive equations. 

Since we work with the brane enclosed in a box by imposing Dirichlet boundary conditions for fields at the isodilatonic surface $r = R$, we have in effect isolated the effective degrees of freedom of the D3-brane at the scale set by $R$. Let us review why this is so: to identify the degrees of freedom associated with the effective dynamics of a D3-brane (or indeed any black brane following the blackfold approach), we need to ascertain how the deviations from the exact black brane metric \eqref{hsD3} can affect the spacetime geometry. While one could focus on solving the exact field equations retaining the asymptotically flat region, the general philosophy of the blackfold approach, or indeed the membrane paradigm, suggests that we should be able to view the dynamics of the black object in terms of an effective dynamics of a `stretched horizon'. While the discussion of the membrane paradigm takes a fiducial timelike hypersurface in the vicinity of the event horizon as the location of the stretched horizon, for the purposes of identifying the fluctuating modes at other scales, it is natural to place the stretched horizon further away. In extreme cases, as in the fluid/gravity correspondence or the blackfold approach, we can indeed take the surface where we project the dynamics all the way to infinity. In effect the dynamics induced on the surface $r = R$ should be the relevant one for understanding the behaviour of the local physics at that scale.

We should also note that, at the level of fluid dynamics, the descriptions at different radial cutoffs are related to each other. Following the discussion of \cite{Brattan:2011my} we can relate the dynamics of the fluid at $r= R_2$ to that at $r=R_1 > R_2$; we view the former as a `dressed fluid'. What this means is the following: the fluid on the surface $r=R_1$ can be subjected to external forces, for example we can turn on a background gravitational potential. While we usually consider background sources that are independent of what the local degrees of freedom are doing, it is nevertheless possible to consider external sources which depend on the local configuration. In the fluid dynamical context  we can subject the fluid at $r=R_1$ to a background gravitational field by changing the metric from $h_{ab} = \eta_{ab}$ to a different metric which depends on the fluid velocity and energy density $h_{ab}\left[u^a, \varepsilon_Q\right]$. In the long-wavelength limit,  absorbing the sources into the dynamical degrees of freedom through a field redefinition allows us to recover the description in terms of a different fluid on a background with $h_{ab} = \eta_{ab}$. This redefined or dressed fluid is indeed nothing but the fluid that lives on the surface $r = R_2$. Explicit maps and field redefinitions for the asymptotically AdS black branes were worked out in \cite{Brattan:2011my} and it is clear that a similar story will apply here. Further details of why the imposition of a rigid cutoff allows us to recover the effective degrees of freedom will be discussed in \cite{Emparan:2013fk}.

Let us now turn to an analysis of the fluid at various scales. The fluid dynamical system we derived depends on two length scales set by $r_+$ and $R$. We can view the former as the overall scale for local energy density or temperature and the latter accounting for location of the Dirichlet boundary. Note that the local temperature at $r=R$ for the black D3-brane is given by working out the period of the Euclidean time circle in \eqref{5dsoln} in the rescaled variables. One has 
\begin{equation}
T = \frac{1}{\pi\, r_+}\frac{\Delta_-(r_+)^{1/4}}{\left(\Delta_{+R}\Delta_{-R}^{-1/2}\right)^{1/2}}=
\frac{1}{\pi\, \ell_Q}\, \frac{\bE^{1/4}(\bE + \bp - \bE \, \bp)^{1/4}(1-\bE)^{1/8}}{\bp^{1/2}}\,,
\label{localT}
\end{equation}	
where we have chosen to express the result in terms of the non-extremality parameter $\bE$, the dimensionless cutoff scale $\bp$, and the charge radius $\ell_Q$. We could use this local temperature as the dynamical variable and express the hydrodynamic data $\{\varepsilon_Q, P_Q, \eta_Q, \zeta_Q\}$ in terms of it, converting from $r_+$ and $R$  using \eqref{rpmRbEbp}. It is however convenient to leave the temperature dependence implicit to avoid cluttering up the formulae.

The entropy density is
\begin{equation}
 s = \frac{2\,\pi}{\kappa_5^2} \, r_+^5\, \frac{\Delta_-(r_+)^\frac{3}{4}}{\Delta_{-R}^\frac{3}{4}}
 = \frac{2\,\pi}{\kappa_5^2} \, \frac{\ell_Q^5}{(1-\bE)^{5/8}}\, \left( \frac{\bE}{\bE + \bp - \bE \, \bp} \right)^{3/4}
 \,.
\label{sdensity}
\end{equation}	
To see this, note that the total entropy is measured as the area of the horizon in Planck units and is thus independent of $R$. However, the entropy density $s$ gets a red-shift factor since the spatial directions are scaled by a factor of $\Delta_-(r_+)/\Delta_{-R}$. We have also chosen to work directly in five dimensions and note that the area can be read-off from the Einstein frame metric \eqref{fivesolef}. Observe that we recover the correct thermodynamic relation
\beq
\varepsilon_Q+P_Q=T s\,.
\eeq

It is useful to consider various limits, which is easily done by looking at the values of the parameters $\bp$ and $\bE$ at the edges of their respective domains. The temperature \eqref{localT} vanishes at extremality ($\bE=0$) for fixed $\bp$ and diverges in the limit to the horizon, $\bp \to 0$, as it must.  The following table lists the interesting limits of ${\mathfrak s}$ and ${\mathfrak v}$:
\begin{center}
\begin{tabular}{|c||c|c|}
\hline 
 ${\mathfrak s}$& $\bp \to 0$ & $\bp \to 1$\\
  \hline\hline
 $\bE \to 0$ & $0$ & $6 \, R^4 \, r_+^{12}$ \\ \hline
 $\bE \to 1$ & $5 \, r_+^{16}$ & $5 \, R^4 \, r_+^{12}$ \\ \hline
\end{tabular}
\hspace{2cm}
\begin{tabular}{|c||c|c|}
\hline 
 ${\mathfrak v}$& $\bp \to 0$ & $\bp \to 1$\\
  \hline\hline
 $\bE \to 0$ & $0$ & $-2\, R^8\, r_+^{12}$ \\ \hline
 $\bE \to 1$ & $ -2 \, r_+^{20}$ &  $ R^8\,r_+^{12}$ \\ \hline
\end{tabular}
\end{center}
which enter in our physical variables (thermodynamics and transport). Note that since $R\to \infty$ as $\bp \to 1$, we keep the dependence on $R$ explicit in the asymptotic values (recall that $1-\bp  = \frac{r_+^4}{R^4}$).  Using these results it is immediate to infer that  the hydrodynamic parameters behave as described in these tables:
\begin{center}
	\begin{tabular}{|c||c|c|}
	\hline 
	 $v_s^2$& $\bp \to 0$ & $\bp \to 1$\\
	 [1ex] \hline\hline\\[-3ex]
	 $\bE \to 0$ & $\frac{1}{3}$ & $\frac{1}{3}$ \\  [1ex] \hline \\ [-3ex] 
	 $\bE \to 1$ & $\frac{2}{5\,\bp}$ & $-\frac{1}{5}$ \\[1ex] \hline
	\end{tabular} 
	\\ 
\vspace{2mm}
\begin{tabular}{|c||c|c|}
\hline 
 $2\,\kappa_5^2\, \eta_Q$& $\bp \to 0$ & $\bp \to 1$\\
 [1ex] \hline\hline\\[-3ex]
 $\bE \to 0$ & $\ell_Q^5 \left(\bE/\bp\right)^\frac{3}{4}$ & 
 $\ell_Q^5 \, \bE^\frac{3}{4}$
 \\ [1ex] \hline \\ [-3ex] 
 $\bE \to 1$ & 
 $\ell_Q^5 \, (1-\bE)^{-\frac{5}{8}}$ 
 & 
 $\ell_Q^5 \, (1-\bE)^{-\frac{5}{8}}$ 
  \\[1ex] \hline
\end{tabular}
\hspace{12mm}
\begin{tabular}{|c||c|c|}
\hline 
 $2\,\kappa_5^2\,\zeta_Q$& $\bp \to 0$ & $\bp \to 1$\\
 [1ex] \hline\hline\\[-3ex]
 $\bE \to 0$ & $\frac{20}{27}\, \ell_Q^5\,\bE^\frac{11}{4}\,\bp^{-\frac{3}{4}}$ & $\frac{20}{27}\, \ell_Q^5\,\bE^\frac{11}{4}$ \\  [1ex] \hline \\ [-3ex] 
 $\bE \to 1$ & 
   $ \frac{16}{15}\, \ell_Q^5 \, (1-\bE)^{-\frac{5}{8}}$
  &  
   $ \frac{16}{15}\, \ell_Q^5 \, (1-\bE)^{-\frac{5}{8}}$
   \\ [1ex]\hline
\end{tabular}
\end{center}
The limiting values for small $\bE$ and $\bp$ are given assuming\footnote{
Note that the value of the hydrodydynamic parameters depends on the order of limits as we take both $\bE $ and $\bp$ to zero. For instance taking $\bp \to 0$ first before cooling down the system to extremality reveals a finite shear viscosity, $\eta_Q = \frac{\ell_Q^5}{2\,\kappa_5^2}$, but this is unphysical since we are sitting on the event horizon. We will return to the extremal D3-brane below.
} $\bE/\bp \to 0$, so that we retain some distance between the horizon and the cutoff surface.

\begin{figure}[h!]
\begin{center}
\includegraphics[width=6in]{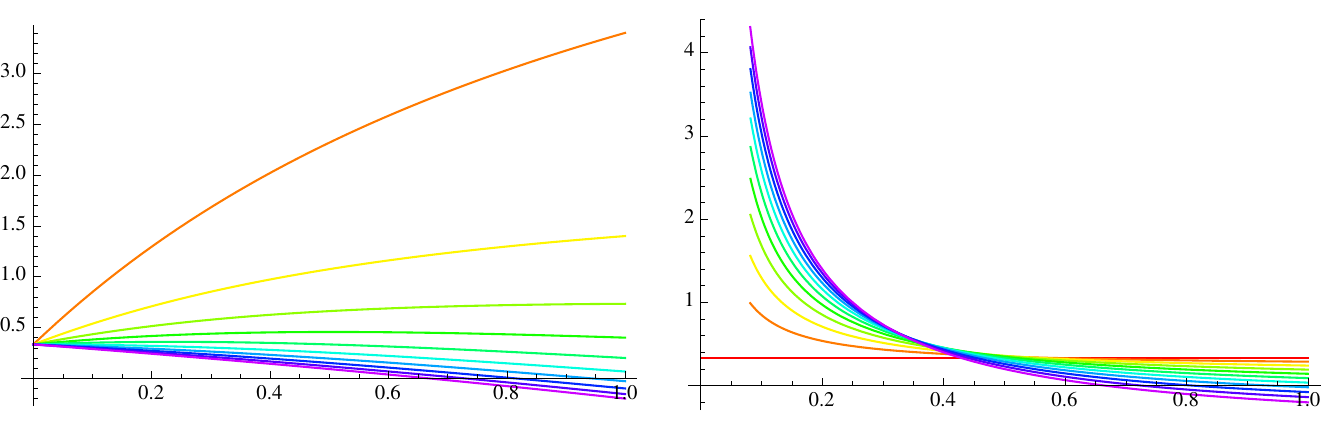}\\
\vspace{1cm}
\includegraphics[width=6in]{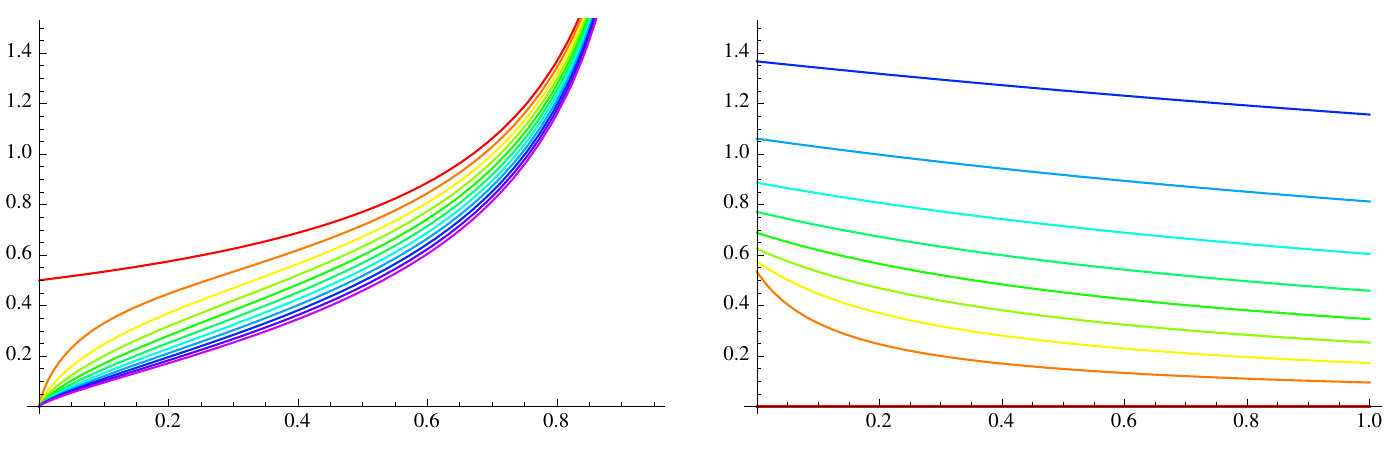}\\
\vspace{1cm}
\includegraphics[width=6in]{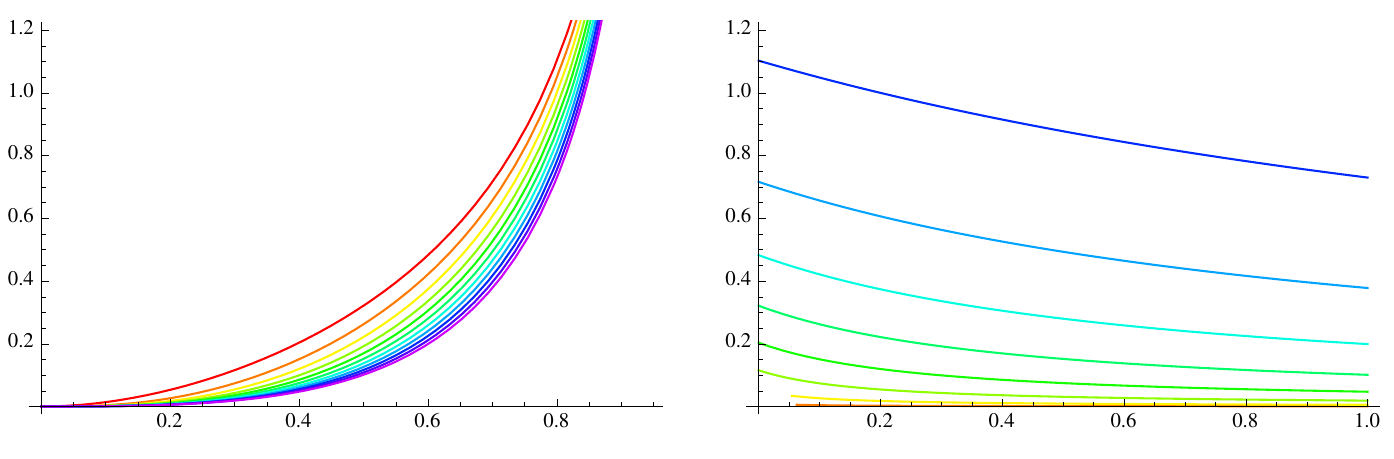}
\begin{picture}(0,0)
\setlength{\unitlength}{1cm}
\put (0,0) {$\bp$}
\put (-7.7,5) {$\zeta_Q$}
\put (-8,0) {$\bE$}
\put (-15.7,5) {$\zeta_Q$}
\put (0,6) {$\bp$}
\put (-7.7,11) {$\eta_Q$}
\put (-8,6) {$\bE$}
\put (-15.7,11) {$\eta_Q$}
\put (0,12) {$\bp$}
\put (-7.7,17) {$v_s^2$}
\put (-8,12) {$\bE$}
\put (-15.7,17) {$v_s^2$}
\put(-13,13.5){$\downarrow \,\bp$}
\put(-13,8){$\downarrow \,\bp$}
\put(-13,1){$\downarrow \,\bp$}
\put(-6.7,13.2){$\uparrow \,\bE$}
\put(-6.5,9.5){$\uparrow \,\bE$}
\put(-6.5,2){$\uparrow \,\bE$}
\end{picture}
\caption{
Plots of the hydrodynamic data as functions of $\bE$ and $\bp$ illustrating the behaviour as we change the nature of the black brane and the location of the cutoff surface. 
The conventions are as in \fig{f:epsQPQ} and the viscosities are plotted in units of $\ell_Q^5/\kappa_5^2$, and the speed of sound in units of speed of light. Once again the finite value of shear viscosity at extremality on the horizon $\bp =0$ is merely due to an order of limits.
} 
\label{f:hydropE}
\end{center}
\end{figure}
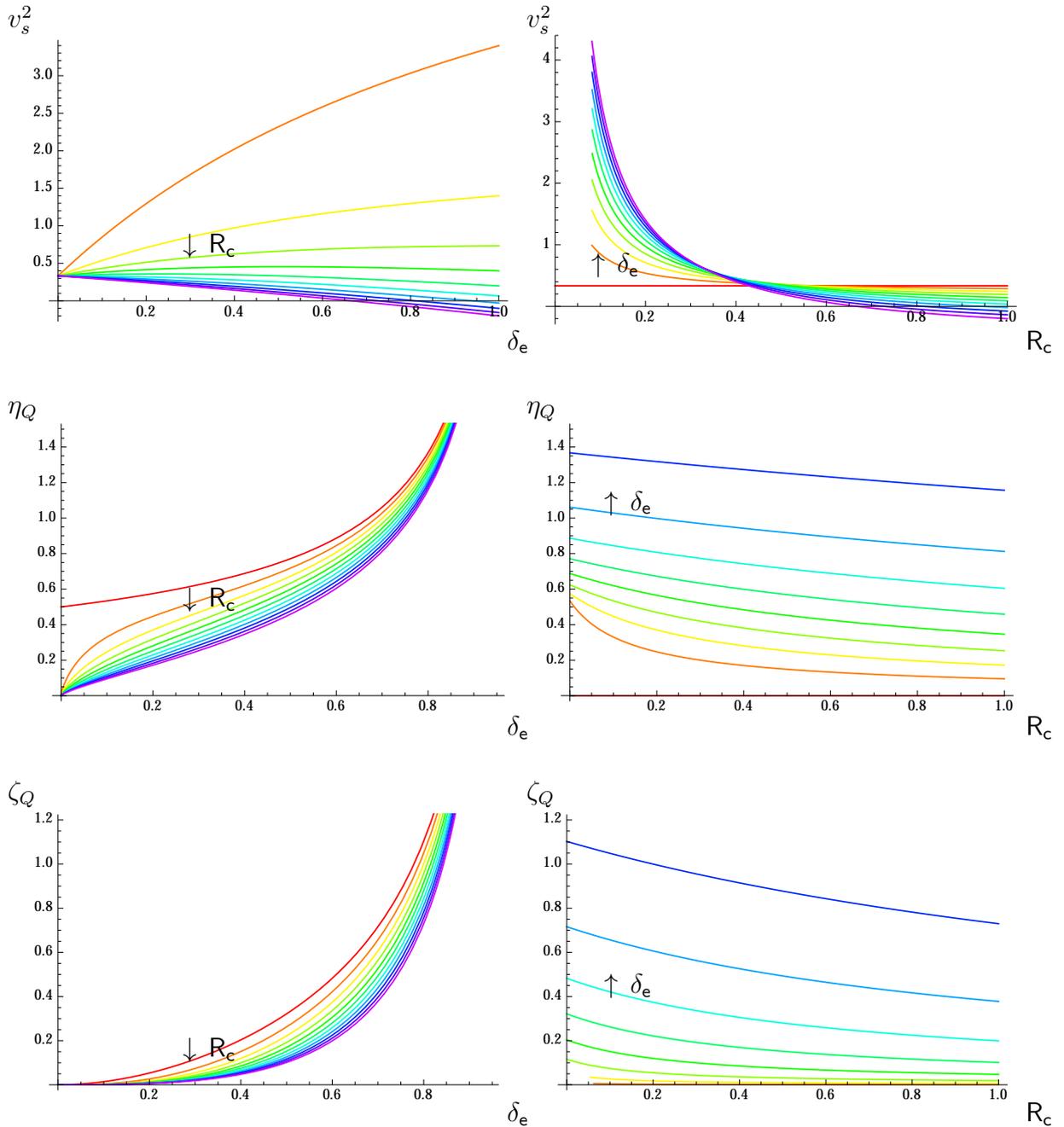

To visualize the behaviour of the fluid we plot the transport coefficients and speed of sound  as functions of $\bE$ and $\bp$ in \fig{f:hydropE}. There are a few salient features that we can recognize. Let us start with thermodynamics that is captured by the speed of sound. For instance, we see that:
\begin{itemize}
\item[(i).] 
For fixed charge, $v_s^2$ increases as $\bp$ decreases.
One way to think of this intuitively is that imposing Dirichlet boundary conditions closer to the horizon leaves less `room' for fluctuations in remaining part of the geometry (between this cutoff surface and the horizon), so it effectively rigidifies the system (this effect was analyzed for neutral black branes in \cite{Emparan:2012be}).  In fact,
$v_s^2 \to \infty$ as we move the cutoff surface all the way to the horizon (except for the special case when the system is strictly at extremality, which we will revisit below). This is the Rindler fluid limit discussed earlier in \cite{Bredberg:2011jq} and indeed we expect the sound mode to freeze out in the near horizon limit, driving the system non-relativistic.
\item[(ii).] 
When the cutoff surface is sufficiently close to the horizon ($\bp \lesssim 1/3$)
the speed of sound is monotonically decreasing with the charge of the D3-brane; excitations propagate fastest in the neutral brane geometry. 
This is because the higher charge causes   deeper throat, so conversely, less charge again effectively rigidifies the system.  On the other hand, for asymptotically placed cutoff surface, the speed of sound actually increases with increasing charge.  Here the above-mentioned throat elongation is no longer the dominant effect.
\item[(iii).] In the limit $\bE\to 0$ the speed $v_s^2$ tends, at all positions of the cutoff $\bp>0$, to the same value. Although this might give the impression that there are sound waves at extremality, this is actually not the case --- at strict extremality there is no fluid. We will discuss this point in more detail below.
\item[(iv).] Finally, note that $v_s^2$ is positive definite only for the cutoff surface being sufficiently near the horizon, or for sufficient amount of charge on the branes. The fact that $v_s^2 <0$ outside this regime is indicative of the Gregory-Laflamme instability of the black branes. As expected the instability is strongest for neutral branes, cf., \cite{Gath:2013qya}. The behaviour as a function of the radial cutoff is qualitatively similar to  that found in the case of black strings \cite{Emparan:2012be}. 
\end{itemize} 

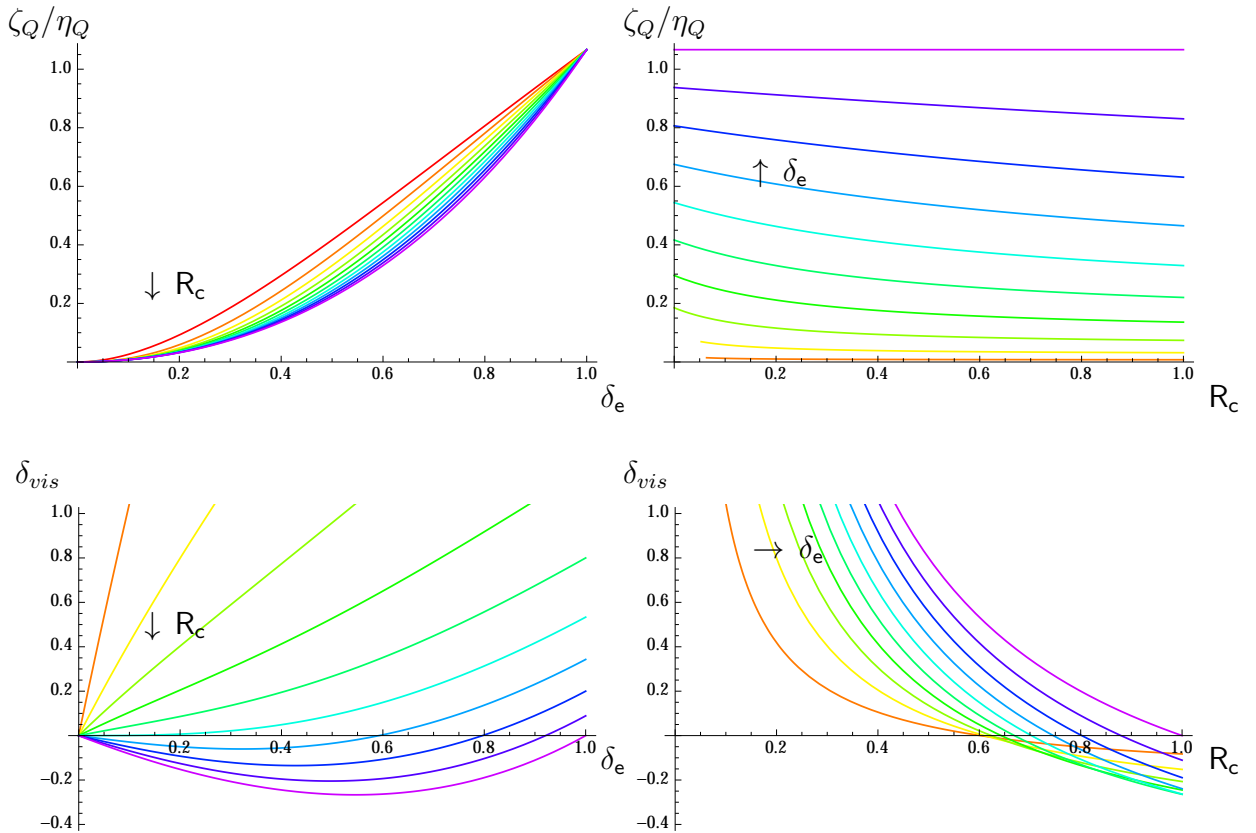
\begin{figure}[h!]
\begin{center}
\includegraphics[width=6in]{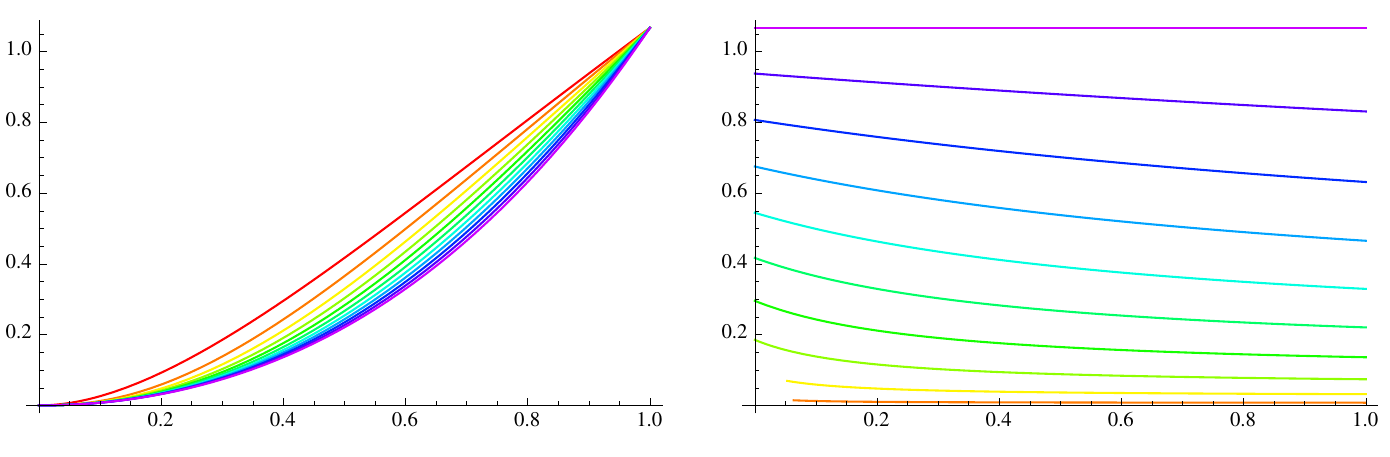}\\
\vspace{1cm}
\includegraphics[width=6in]{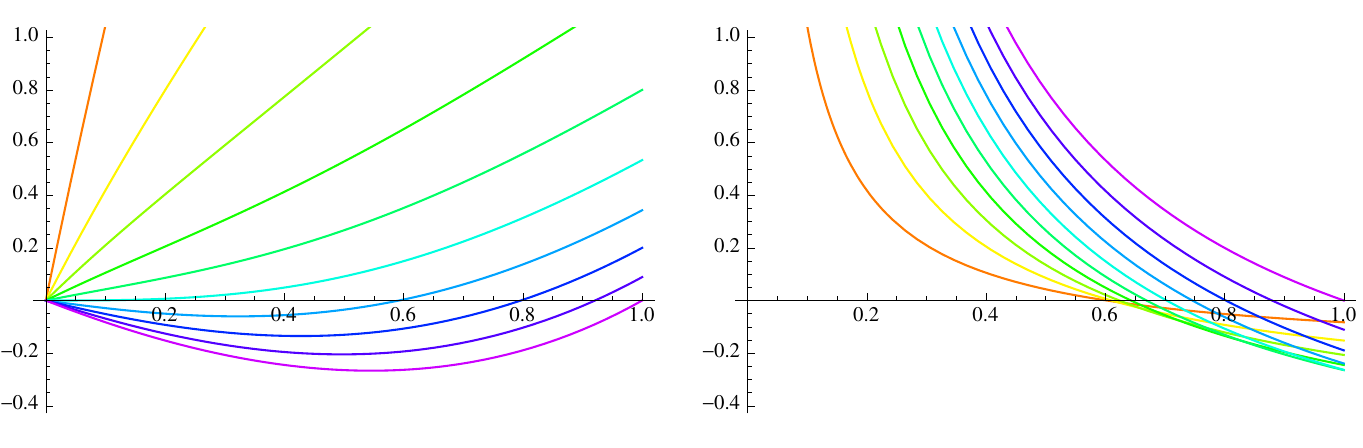}
\begin{picture}(0,0)
\setlength{\unitlength}{1cm}
\put (0,1.2) {$\bp$}
\put (-7.7,5) {$\delta_{vis}$}
\put (-8,1.2) {$\bE$}
\put (-15.7,5) {$\delta_{vis}$}
\put (0,6) {$\bp$}
\put (-7.7,11) {$\zeta_Q/\eta_Q$}
\put (-8,6) {$\bE$}
\put (-15.8,11) {$\zeta_Q/\eta_Q$}
\put(-14,7.5){$\downarrow \,\bp$}
\put(-14,3){$\downarrow \,\bp$}
\put(-6,9){$\uparrow \,\bE$}
\put(-6,4){$\rightarrow \,\bE$}
\end{picture}
\caption{
Plots of the ratio of the viscosities of the black D3-brane as functions of $\bE$ and $\bp$. The conventions are as in \fig{f:epsQPQ}.
In the top panels we show the behaviour of $\zeta_Q/\eta_Q$ and in the bottom panels we plot $\delta_{vis}  = \frac{\zeta_Q}{\eta_Q} - 2\left(\frac{1}{3} - v_s^2\right)$ to illustrate that the bulk viscosity bound is violated.} 
\label{f:ezratio}
\end{center}
\end{figure}

Having discussed the behaviour of the speed of sound, let us consider the transport coefficients. Using the expressions \eqref{etaD3} and \eqref{zetaD3} we can see that:
\begin{itemize}
\item[(a).] The shear viscosity is independent of the cutoff surface $R$ for neutral branes (see below); this was verified originally in \cite{Emparan:2012be}. This is no longer the case for charged branes. 
In general $\eta_Q$  increases as we move the cutoff closer to the horizon. Nevertheless the ratio $\eta_Q/s$, with $s$ given in \eqref{sdensity}, is constant and, as one expects for two derivative gravity theories, saturates the famous KSS bound \cite{Kovtun:2004de}, $\eta_Q/s = \frac{1}{4\pi}$.

Furthermore, $\eta_Q$ depends on the deviation from extremality $\bE$; for fixed $\bp$ it decreases as we approach extremality, with $\eta_Q \to 0$ as $\bE \to 0$ (which it must do since $s$ decreases with charge).
\item[(b).] Similarly, the bulk viscosity is independent of the cutoff for the neutral brane $\bE =1$, and vanishes as $\bE \to 0$ --- which, as we discuss below, does actually reflect a property of the system to leading near-extremal order, and not just at strict extremality. The bulk viscosity also decreases when we increase the charge on the brane and grows as we move the cutoff surface to the horizon.
\item[(c).] It is easy to see that ratio of viscosities violates the bound conjectured in \cite{Buchel:2007mf}, viz., the inequality $\frac{\zeta_Q}{\eta_Q} \geq 2\, \left(\frac{1}{3} - v_s^2\right)$ fails to hold in general. This happens for a brane carrying sufficient amount of charge or for sufficiently large radial cutoffs. We illustrate this behaviour in \fig{f:ezratio}.
\end{itemize}

\begin{figure}[h!]
\begin{center}
\includegraphics[width=6in]{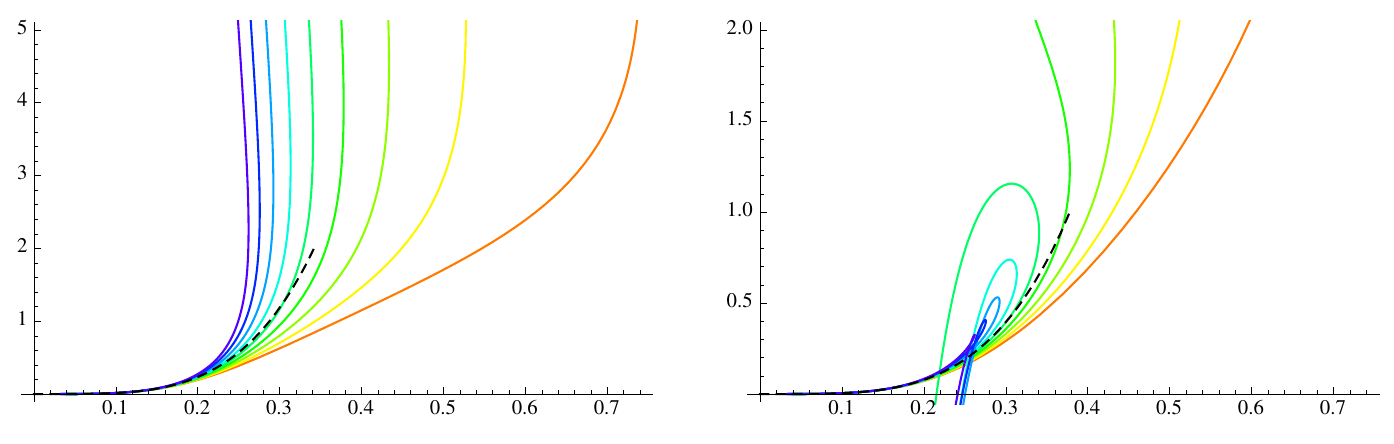}\\
\vspace{1cm}
\includegraphics[width=6in]{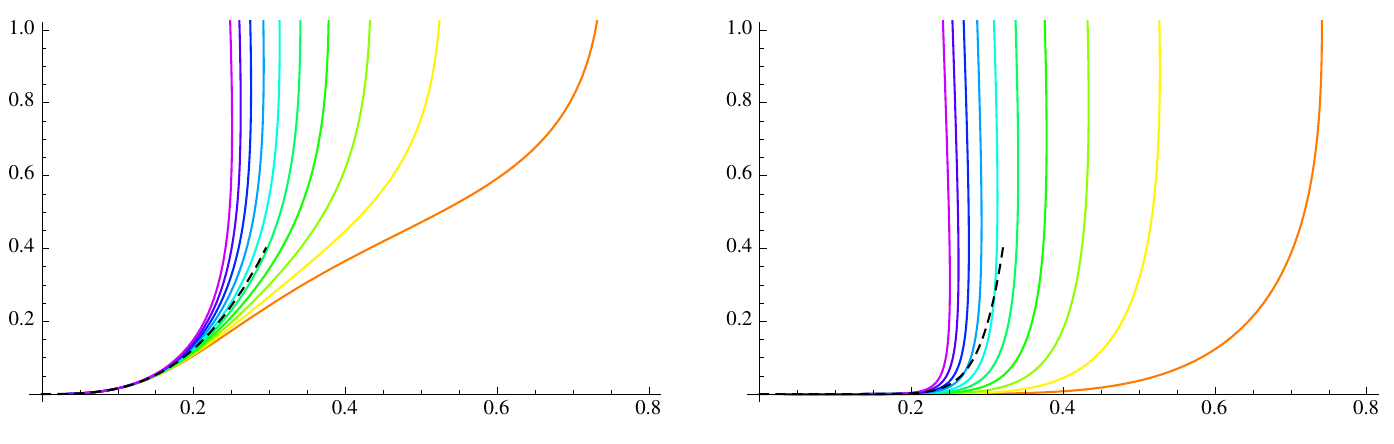}
\begin{picture}(0,0)
\setlength{\unitlength}{1cm}
\put (0,0.2) {$T$}
\put (-7.7,5) {$\zeta_Q$}
\put (-8,0.2) {$T$}
\put (-15.7,5) {$\eta_Q$}
\put (0,6) {$T$}
\put (-7.7,10.8) {$P_Q$}
\put (-8,6) {$T$}
\put (-15.6,10.8) {$\varepsilon_Q$}
\put(-12,10){$\leftarrow \,\bp$}
\put(-12,4){$\leftarrow \,\bp$}
\put(-4,10){$\leftarrow \,\bp$}
\put(-4,4){$\leftarrow \,\bp$}
\end{picture}
\caption{
Plots of hydrodynamic parameters as function of the local temperature $T$ on the cutoff surface. To guide the eye we show the near-extremal behaviour by the dashed (black) line. The colour coding conventions for $\bp$ are as in \fig{f:epsQPQ}.} 
\label{f:hydroT}
\end{center}
\end{figure}
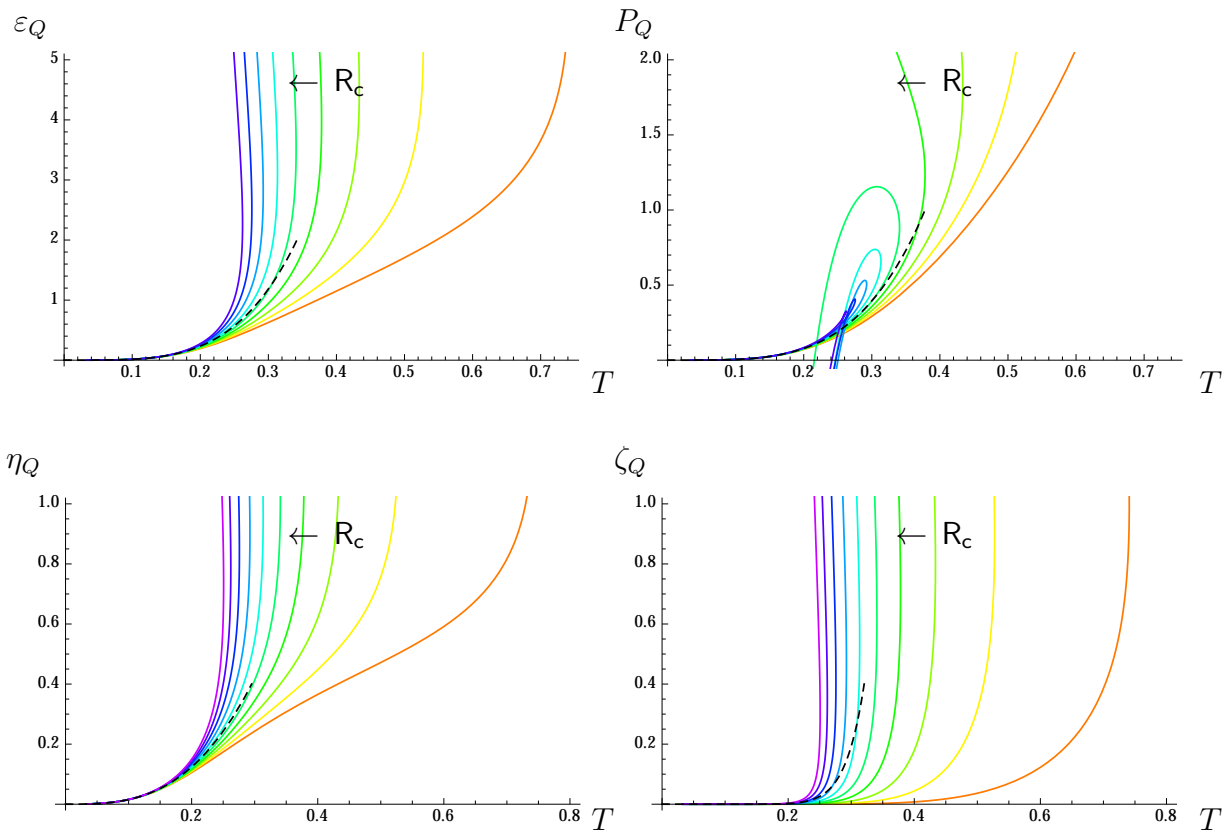

It is also interesting to the visualize the hydrodynamic parameters directly in terms of the local temperature. We display the behaviour of $\{\varepsilon_Q, P_Q, \eta_Q, \zeta_Q\}$ as a function of the local temperature $T$ \eqref{localT} in \fig{f:hydroT}.
The  curves for different values of $\bp$ stay bunched together for small $T$, which corresponds to the near-extremal limit -- we discuss this more in detail below. 
In the plot for $P_Q(T)$ we observe a swallow-tail behaviour for $R \to \infty$, while the other quantities, like $\varepsilon_Q(T)$, also exhibit double-valuedness at large $R$. This is indeed a familiar effect from charged black holes in the canonical ensemble --- the black D3-brane, when compactified on a 3-torus, yields a charged dilatonic black hole. For these systems, it is known that at fixed charge, and for fixed temperature below a maximum (critical) value, there are two possible black holes each with different mass. At the critical value the two solutions merge and at higher temperatures there is no black hole with those values of $T$ and $Q$. In our case, the black hole mass corresponds to the energy density of the black brane --- and thus we see, at fixed and large enough $\bp$, a double-valuedness of $\varepsilon_Q$, up to a maximum temperature. The free energy of the black hole is (minus) the pressure $P_Q$ of the black brane: this free energy possesses at large $\bp$ the kind of swallow-tail structure that we observe in the plots.\\

Let us finally examine two interesting limiting cases in some detail. Firstly, note that the constitutive relations for the fluid reduce to those of the neutral brane when we set $\ell_Q =0$. Note that the limit $\ell_Q \to 0$ should be taken with care when working with the parameterization in terms of $\bE$ and $\bp$. The correct limiting behaviour is the one where we keep $r_+ = \ell_Q \, (1-\bE)^{-\frac{1}{8}}$ fixed as we take $\ell_Q$ to zero, ensuring thereby the we have a black hole with a non-vanishing horizon size while we detune the charge. As noted in the points above, the neutral brane has no variation of the transport coefficients as we move the cutoff surface; i.e., $\eta_{Q=0}$ and $\zeta_{Q=0}$ are independent of $R$. These have been extensively studied in the past and a discussion of the hydrodynamic properties of neutral branes is provided in \cite{Emparan:2012be}.

The other limit is the near-extremal brane. Expanding in $\bE\ll 1$ we get
\begin{align}
\varepsilon_Q&=\frac{\ell_Q^4}{2\,\kappa_5^2}\left(\frac{3\,\bE}{\bp}+\frac{17\,\bp -9}{4\,\bp^2}\,\bE^2+{\cal O}(\bE^3)\right)\,,\notag\\
P_Q&=\frac{\ell_Q^4}{2\,\kappa_5^2}\left(\frac{\bE}{\bp}+\frac{1-\,\bp}{4\,\bp^2}\,\bE^2+{\cal O}(\bE^3)\right)\,.
\end{align}
In the extremal limit both energy density and pressure vanish: this is due to our choice of counter-terms, which have been chosen so as to subtract all non-hydrodynamic contributions to the stress-energy. Thus in the extremal limit the fluid density vanishes and hydrodynamics does not apply. The leading order near-extremal contribution is, however, of quite some interest to us. Using the temperature \eqref{localT} we can write 
\begin{align}
\varepsilon_Q&= \frac{\ell_Q^8}{2\,\kappa_5^2}\; 3\, (\pi\,T)^4 \left(1+{\cal O}(\bE)\right)\,,\notag\\
P_Q&=\frac{\varepsilon_Q}{3}\left(1+{\cal O}(\bE)\right)\,.
\label{n4eos}
\end{align}
This is a conformal fluid: $\varepsilon_Q=3 P_Q\propto T^4$. Then the sound speed in the leading near-extremal fluid is $v_s^2=1/3$, \textit{independently of $\bp$}. This value indicates that the long-wavelength modes obey scale-invariant dispersions. The energy density $\varepsilon_Q$ of these near-extremal excitations does depend on $\bp$, but their dependence is fully dictated by that of $T$, and in turn the dependence of the latter on $\bp$ simply corresponds to the redshift factor at the location of the cutoff surface. So all the $\bp$-dependence of the fluid, at the leading near-extremal order, is trivial. This universal behaviour for low temperatures can be seen clearly in \fig{f:hydroT}.

At next-to-leading near-extremality the fluid, however, does deviate away from this behaviour and depends non-trivially on $\bp$. We can see this in 
\begin{equation}
v_s^2 =\left(\frac{\partial P_Q}{\partial\varepsilon_Q}\right)_R= \frac{1}{3} + \left(\frac{2}{3\,\bp} - \frac{10}{9} \right)\bE + {\cal O}\left(\bE^2\right)\,.
\label{vs2}
\end{equation}	
Recall that this velocity is calculated on an isodilatonic surface at constant $R$.
Let us emphasize that, even if $v_s^2$ remains finite as $\bE\to 0$, this result is \textit{not} the speed of sound of an extremal brane, but rather of a black brane with thermal excitations infinitesimally away from extremality.

 Turning to the viscosities, to leading near-extremal order we have
\begin{equation}
\eta_Q=\frac{\ell_Q^8}{2\,\kappa_5^2}\, (\pi\,T)^3\left(1+{\cal O}\left(\bE\right)\right)\,,
\end{equation}
while $\zeta_Q$ vanishes with a higher power of $\bE$. Thus we find that the hydrodynamic transport coefficients are those of a conformal fluid without any non-trivial dependence on the position of the cutoff surface $\bp$.

These results for the leading near-extremal fluid are of course familiar: they are the same as one finds for the hydrodynamics of AdS black branes. We can now interpret the independence from $\bp$ of the leading near-extremal hydrodynamics of black D3-branes.\footnote{As we note around \eqref{d3charge} our conventions are that $\kappa_5^2 = \frac{8\pi\, G^{(10)}_N}{\pi^3}$ with the factor of $\pi^3$ coming from the volume of the ${\bf S}^5$. It then follows that $\ell_Q^8/\kappa_5^2 = \frac{N^2}{4\pi^2}$ where $N$ is the total number of D3-branes, recovering thus in \eqref{n4eos} the standard result for the $SU(N)$, ${\cal N}=4$ Super Yang-Mills theory at strong coupling.} Recall that the decoupling limit of the black D3-brane, which isolates the dynamics of AdS black branes, retains precisely such infinitesimal deviations away from extremality --- these are the energies of the excitations that decouple from the asymptotically flat region beyond the throat. What we have found is a manifestation of this decoupling: when we keep the asymptotically flat region and measure the hydrodynamics on a cutoff surface beyond the throat region, then this hydrodynamics is that of the modes inside the throat, unaffected by the presence of the exterior region. The cutoff surface only introduces a trivial redshift corresponding to observing the conformal fluid from a distance.

We can then say that most of the action for the low energy dynamics takes place in the AdS throat as we approach $\bE \to 0$. Examining the behaviour of the system for $\bE \ll 1$ we find that the modifications to the effective fluid dynamics are small away from the throat region. The dynamics is controlled by the throat, and the exterior of the AdS region simply serves to connect up to the asymptotically flat spacetime. The modifications of the effective dynamics that the exterior region introduces are only significant for excitation energies beyond the leading near-extremal order. We have verified the trivial dependence on $\bp$ only up to first-order hydrodynamics, but the decoupling argument indicates that we should expect it to persist to all orders. 

Thus, we have found in a very explicit way how the hydrodynamics of the AdS black branes, i.e., the fluid/gravity correspondence of \cite{Bhattacharyya:2008jc}, emerges out of the blackfold-type hydrodynamics of black D3-branes.

\section{Discussion}
\label{s:discuss}

The primary result of our analysis is that a black D3-brane geometry with rigid Dirichlet boundary conditions imposed on an isodilatonic hypersurface $r = R$ obeys the equations of a relativistic dissipative fluid. To be precise, as in the fluid/gravity correspondence \cite{Hubeny:2011hd}, we have shown that the dynamics of gravity as described by Einstein's equations between the horizon and the cutoff surface maps onto the conservation equation for the stress tensor $\nabla_a T^{ab} =0$. The constitutive relations for the fluid system are given in \eqref{zeroST}, \eqref{etaD3} and \eqref{zetaD3} respectively.

We should emphasize at this stage that our discussion has been restricted to fluctuations of the D3-brane (with the aforementioned boundary conditions) along its world-volume.  As is well known, the D3-brane   can also be excited in the transverse directions; the dynamics of these modes is captured by a different physical system, viz., an elastic solid \cite{Armas:2011uf,Camps:2012hw}. At leading order in the long-wavelength approximation the intrinsic dynamics decouples from the extrinsic one. We have exploited this decoupling to focus on the former exclusively. If we were to consider higher order effects in the hydrodynamic description, we would necessarily have to worry about the coupling between these two types of modes. We should mention here that the extrinsic modes freeze out as we take the cutoff surface down the AdS throat in the case of near-extremal branes.  

By focussing on the intrinsic dynamics we have been able to ascertain the relation between the different approaches to the map between hydrodynamics and gravity. As we previewed in \sec{s:intro}, by exploiting the freedom to cut off the geometry on a fiducial timelike surface $r=R$ we have been able to isolate the dynamical system operational at a given scale around the black brane horizon. Effectively, we are using this surface as a stretched horizon.  Since the system has two dimensionless parameters, the horizon size $r_+$ and the cutoff scale $R$ (measured in units of the fixed charge length $\ell_Q$), we have the ability to tune into the different regions of the geometry. As a result we see that non-relativistic membrane fluid description holds true in the ultra near-horizon Rindler geometry, while the fluid/gravity conformal fluid description starts to apply at the bottom of the AdS throat (for near extremal branes). Generically however as $R \gg r_+ > \ell_Q$ we don't have an AdS region and the dynamics is best described in the language of the blackfold effective theory. As a consequence we see that the inclusion relation \eqref{inclusionrel} holds true in our example. The D3-brane geometry provides the simplest setting in which we can embed all these different fluid dynamical systems as the effective low energy dynamics.

One surprising result of our analysis is the fact that viscosities of the black D3-brane vary non-trivially as a function of the radial cutoff. In previous analysis for the neutral brane \cite{Emparan:2012be} the bulk and shear viscosities did not change with the position of the radial cutoff. Similarly the behaviour of shear and bulk viscosities in the case of AdS-black branes examined in \cite{Brattan:2011my} also showed no dependence on the cutoff location for the transport parameters. Despite this novel feature, it nevertheless remains true that the ratio of shear-viscosity to entropy density saturates the KSS bound. However, the ratio of bulk viscosity to shear viscosity violates the so-called Buchel bound in certain regions of parameter space. The other fact which we have elaborated on in some detail in \sec{s:physics} is that the hydrodynamic behaviour to leading order away from extremality is fully controlled by the throat and not affected by the region exterior to it. We interpret this as a manifestation of the decoupling of these throat excitations from the asymptotically flat exterior. In effect the near-extremal geometry acts as an {\em attractor} -- the enhanced symmetry manifested in this regime controls a large part of the parameter space.

It is clear that our analysis can be generalized to various other branes in string/M-theory. Obvious extensions to the non-dilatonic M2 and M5 branes will recover very similar behaviour albeit in $2+1$ and $5+1$ dimensions respectively for the world-volume dynamics. The non-dilatonic branes can also be easily studied following our  analysis; in these cases it would be interesting to examine the cutoff dependence of the transport coefficients.  In addition to these obvious extensions, one could also consider varying the Ramond-Ramond charge carried by the branes along the world-volume directions. For example, one could consider a D0-D4 system wherein the D0 brane density on the world-volume of the D4-branes fluctuates. Such systems have not hitherto been studied in the long-wavelength effective description and it would be interesting to figure out the low energy effective dynamics with higher form potentials.

As  noted in the course of the computation, we have not considered the most general stationary black  D3-brane solution as the starting point of our long-wavelength perturbation theory. In principle one can also let the branes carry
 angular momentum in the transverse ${\mathbb R}^6$. In the world-volume theory we will then encounter Maxwell fields and the resulting dynamical system will be that of a charged fluid carrying $U(1)^3$ charges. In the near-extremal limit the geometry down the AdS throat then reduces to that of an AdS-Reissner-Nordstr\"om black brane, which has proven to be a rich ground for the study of hydrodynamics, revealing e.g.\ the presence of anomalous transport coefficients arising from quantum anomalies \cite{Erdmenger:2008rm,Banerjee:2008th}.

Finally, while we have asserted that the D3-brane geometry allows one to interpolate between the membrane paradigm, fluid/gravity, and the blackfold approaches, we have relied mostly  on the recent constructions of the membrane fluid in terms of the Rindler hydrodynamics to identify the membrane description. The identification of the conformal fluid encountered in the AdS fluid/gravity context and the blackfold fluid on the other hand is quite clean and unambiguous.  The precise connection between the non-relativistic fluid  dual to the Rindler geometry and the original idea as envisaged by Damour \cite{Damour:1982fk} will be explained elsewhere \cite{Emparan:2013fk}. We shall also describe there the general framework  effective dynamics of black branes building on various ideas involved in the fluid/gravity and blackfold approaches.

\acknowledgments 
It is a pleasure to thank Joan Camps and Shiraz Minwalla for useful discussions. We would like thank Stefan Steinfurt for spotting some typos in an earlier version of the paper.
RE and MR would like to thank the organizers of the meeting ``Iberian Strings 2012'' in Bilbao for their hospitality during the initial stages of this project. VH and MR would also like to acknowledge  the hospitality of: KITP during the ``Bits, branes and black holes'' workshop, the University of Amsterdam during the Amsterdam Strings Workshop 2012   during the course of this project. RE, VH and MR would like to thank NORDITA for hospitality during the workshop ``The holographic way:
string theory, gauge theory and black holes". RE was partially supported by MEC FPA2010-20807-C02-02, AGAUR 2009-SGR-168 and CPAN CSD2007-00042 Consolider-Ingenio 2010. 
VH and MR were supported in part by the National Science Foundation under Grant No. NSF PHY11-25915 and by the STFC Consolidated Grant ST/J000426/1.

\appendix

\section{Sphere reductions and effective actions}
\label{s:kkreduce}

In this appendix we quickly outline the general strategy for sphere reductions from $D = d+1+N$ dimensions down to $d +1$ dimension on a round ${\bf S}^N$.

\subsection{The reduction ansatz on ${\bf S}^N$}

Let us consider the class of warped-product spacetimes
\beq\label{redN}
ds^2_{d+1+N} =ds^2_{d+1}+ e^{2\varphi(x)}d\Sigma_N^2
\eeq
where $\varphi$ depends only on the coordinates $x^\mu$ of the $(d+1)$-dimensional factor
\beq
ds^2_{d+1}=g_{\mu\nu}(x)dx^\mu dx^\nu\,.
\eeq
Denoting with an overbar the quantities in the full $(d+1+N)$-dimensional spacetime we have
\beq
\bar R=R+e^{-2\,\varphi}R_{\Sigma_N}-2\,N\,\Box\varphi-N(N+1)\,(\partial\varphi)^2
\eeq
and
\beq
\sqrt{-\bar g}\;\bar R= \text{Vol}({\Sigma_N})\sqrt{-g}\left\{ e^{N\varphi}\left[R+e^{-2\varphi}R_{\Sigma_N}+ N(N-1)(\partial\varphi)^2\right]
-2N\nabla_\mu\left(e^{N\varphi}\partial^\mu\varphi\right)\right\}.
\eeq 

When $\Sigma_N$ is a $S^N$ of unit radius, $R_{\Sigma_N}=N(N-1)$, so
\beq\label{Rred}
\sqrt{-\bar g}\;\bar R=\Omega_N \sqrt{-g}\left\{ e^{N\varphi}\left[R+N(N-1)\left((\partial\varphi)^2+e^{-2\varphi}\right)\right]
-2N\nabla_\mu\left(e^{N\varphi}\partial^\mu\varphi\right)\right\}\,.
\eeq 
This allows us to perform the reduction of the Einstein-Hilbert action. The last term is a total derivative and we will ignore it for now.  Also, the Ricci tensor along the $x^\mu$ directions is
\beq
\bar R_{\mu\nu}=R_{\mu\nu}-N\left(\partial_\mu\varphi \partial_\nu\varphi+\nabla_\mu\partial_\nu\varphi\right)\,.
\eeq

The reduced metric is not in Einstein frame, but instead in what we may refer to, given the absence of dressing of the $(d+1)$-dimensional metric, as the `naked frame'. Note also that the dilaton is not canonically normalized. The naked frame has some slight advantages when we are mostly interested in the equations of motion: one avoids awkward numbers in the conformal factors, and the $(d+1+N)$-dimensional solution is read straightforwardly.  Also, the computation of the quasilocal stress tensor is simple, as we will see later.

Note the unusual dimensional homogeneity: $e^\varphi$ is made to have dimension [length]. One might remedy this by introducing a suitable length scale in the metric ansatze for the radius of the ${\bf S}^N$. However, since this is not necessary we will not do it.

The result presented in \eqref{5deffn} follows from here when we set $N =5$ and further account for the contribution of 
the five-form flux. The latter gives the term proportional to $Q^2$ in the effective action. 

\subsection{Five dimensional Einstein frame for black D3-branes}
\label{s:fived}

The reduction ansatze to the five-dimensional Einstein frame, with canonically normalized scalar, is
\beq
ds^2=e^{2\alpha\,\tilde\varphi}ds^2_5+e^{2\beta\,\tilde\varphi}d\Omega_5^2
\eeq
and the 5-form 
\beq
F_{(5)}= Q\left( e^{8\alpha\,\tilde\varphi}\, \epsilon_{(5)} +
\text{Vol}({\bf S}^5)\right)\,.
\eeq
\begin{equation}
S_5 =\frac{1}{2\,\kappa_5^2} \int\, d^5x\,\sqrt{-g}\, \left(R- \frac{1}{2}\, (\partial
\tilde\varphi)^2 -
V(\tilde\varphi)\right)\,, \qquad
V(\tilde\varphi) = 2\,Q^2\, e^{8\,\alpha\, \tilde\varphi}- 20\, e^{\frac{16}{5}\,
\alpha\, \tilde\varphi}
\label{5defeff}
\end{equation}	
with:
\begin{equation}
\alpha = \frac{1}{4}\, \sqrt{\frac{5}{3}} \,, \qquad \beta = -\frac{3}{5}\,\alpha
\label{abdef}
\end{equation}	
Note that in our conventions $\frac{1}{\kappa_5^2} = \frac{\Omega_5}{8\pi\, G_N^{(10)}}$ with $\Omega_5 = \pi^3$ being the volume of a unit ${\bf S}^5$.

The static D3-brane solution in terms of the these fields is
\begin{equation}
ds_5^2 = r^{\frac{10}{3}} \, \left(-\Delta_+ \,
\Delta_-^{-\frac{1}{2}}\, dt^2 +
\Delta_-^\frac{1}{2}\, d{\bf x}^2 + \frac{dr^2}{\Delta_+\,
\Delta_-}\right) \ ,\qquad
e^{-\frac{3}{5}\,\alpha\,\tilde\varphi} = r
\label{fivesolef}
\end{equation}	
%

\section{Quasi-local stress-energy tensor}
\label{s:qlocalst}

We begin by doing an analysis in $D=d+1+ N$ dimensions, then we will restrict to ten-dimensional spacetime and then describe how the result can be obtained directly in the dimensionally reduced geometry. 

\subsection{Higher-dimensional calculation}

We are interested in computing the quasilocal stress-energy tensor on $(D-1)$-dimensional surfaces with induced metric
\beq\label{indmetric}
\bar h_{\mu\nu}dx^\mu dx^\nu=h_{ab}d\sigma^a d\sigma^b +e^{2\boldsymbol\varphi}d\Omega_{(N)}^2\,.
\eeq
The $d$-dimensional metric $h_{ab}$ and dilaton value on the surface, $\boldsymbol\varphi$, are fixed.
We refer to $\sigma^a$ as the coordinates along world-volume directions. 
 
The quasi-local stress-energy tensor along these directions,
integrated over ${\bf S}^N$, and without any subtraction,\footnote{
The overbars denote that the extrinsic curvatures are computed in the $D$-dimensional geometry; the tilde is to refer to the unsubtracted stress-energy.} is
\beq
\tilde T_{ab}=\frac{\Omega_N}{8\pi\, G_N^{(D)}}\; e^{N\boldsymbol\varphi}\lp\bar\Theta_{ab}-h_{ab}\bar\Theta\rp\,.
\eeq
We are not interested in the parts of this stress-energy tensor that are not associated to any intrinsic dynamics. In particular, it is clear that contributions of `vacuum' type, i.e.,  proportional to $h_{ab}$ and with constant energy density, cannot be associated to hydrodynamic behaviour, which breaks Lorentz invariance along the world-volume and introduces inhomogeneities. 

Let us focus on the case when $h_{ab}$ is a flat metric and the value of $\boldsymbol\varphi$ at the boundary is constant, i.e., independent of $\sigma^a$, $h^{ab}\partial_b\boldsymbol\varphi=0$. We may actually define the surface by $\varphi(x)=\boldsymbol\varphi=\text{const}$. We can then consider a surface in flat $D$-dimensional Minkowski spacetime at the given constant ${\bf S}^N$-radius $e^{\boldsymbol\varphi}$ and with the same metric $h_{ab}$. For this surface, $\bar\Theta_{ab}^{(0)}=0$, and the stress tensor, which comes entirely from the curvature of the sphere,  $\bar\Theta^{(0)}=-Ne^{-\boldsymbol\varphi}$, is vacuum-type and equal to
\beq
T_{ab}^{(M)}=\frac{\Omega_N}{8\pi\, G_N^{(D)}}\; N\, e^{(N-1)\boldsymbol\varphi}\; h_{ab}\,.
\eeq
It is easy to show that, on a surface at asymptotic infinity, the quasilocal stress-energy tensor with Minkowski-space subtraction $T_{ab}=\tilde T_{ab}-T_{ab}^{(M)}$ enters the metric as
\beqa\label{asympmetric}
ds^2_d&=&\lp \eta_{ab}+\frac{16\pi\, G_N^{(D)}}{(N-1)\, \Omega_N}
\lp T_{ab}-\frac{1}{D-2}\, T\, \eta_{ab}\rp\frac{1}{r^{N-1}}\rp d\sigma^a d\sigma^b
\nonumber\\ 
&&
+\lp 1-\frac{16\pi\, G_N^{(D)}}{\Omega_N}\; \frac{1}{D-2}\; \frac{T}{r^{N-1}}\rp
dr^2+ O(r^{-2(N-1)})\,,
\eeqa
while $e^\varphi=r$. Here $T=\eta^{ab}T_{ab}$ and $D=d+N$. The stress-energy tensor can then be easily read from the asymptotic behaviour of the metric.

If we are interested in branes that carry a fixed charge $\mathbf{Q}$, we may want to subtract the stress-energy of the ground state, i.e., the extremal brane, since this will not affect the hydrodynamical behaviour. Typically the extremal brane is a BPS object of tension $\mathbf{Q}$, and being a Lorentz-invariant state it has
\beq
T_{ab}^{(Q)}=-\mathbf{Q}\; h_{ab}=-\frac{\Omega_N}{8\pi\, G_N^{(D)}} \, Q \, h_{ab}\,.
\eeq

Thus, we shall take
\beqa
T_{ab}&=&\tilde T_{ab}-T_{ab}^{(M)}-T_{ab}^{(Q)}\\
&=&\frac{\Omega_N}{8\pi\, G_N^{(D)}}\left[e^{N\boldsymbol\varphi}\lp\bar\Theta_{ab}-h_{ab}\bar\Theta\rp
+\left(Q-Ne^{(N-1)\boldsymbol\varphi}\right)h_{ab}\right]\,.
\eeqa

Note that since the subtracted tensors are $\propto h_{ab}$ with proportionality coefficients that, for fixed ${\bf S}^N$ radius and charge, are constant, they only affect the non-dissipative part of the stress-energy tensor, and not the transport coefficients --- this should be clear since these `ground state' terms do not have any hydrodynamic behaviour.  In particular, the transport coefficients are finite even if the surface lies in the asymptotic region and no subtraction is performed.

\subsection{Reduced formulation: naked frame}

Let us now give a purely $d+1$-dimensional formulation of the stress tensor in  
the naked-frame geometry, \eqref{redN}.

Recall that we are assuming that $\boldsymbol\varphi$  is constant along the boundary, \ie $h^{ab}\partial_b {\boldsymbol\varphi}=0$, and that $h_{ab}$ is a flat metric. This is an important simplifying assumption, which in particular allows us to easily embed the boundary in $D$-dimensional Minkowski space.

For these surfaces, the extrinsic curvature tensor is the same as the $D$-dimensional one along world-volume directions,
\beq
\bar\Theta_{ab}=\Theta_{ab}\,.
\eeq
However, the trace of the extrinsic curvature receives a contribution from the ${\bf S}^N$. Since $\sqrt{-\bar h}=e^{N{\boldsymbol\varphi}}\sqrt{-h}$, we have
\beq
\bar\Theta= -\frac{e^{-N{\boldsymbol\varphi}}}{\sqrt{-h}}n^\mu\partial_\mu\lp e^{N{\boldsymbol\varphi}}\sqrt{-h}\rp
=
\Theta -N\,n^\mu\partial_\mu{\boldsymbol\varphi}\,.
\eeq
Then
\beq\label{Tab}
T_{ab}=\frac{\Omega_N}{8\pi\, G_N^{(D)}}\left[e^{N{\boldsymbol\varphi}}\lp\Theta_{ab}-h_{ab}\Theta\rp
+\left(n^\mu\partial_\mu e^{N{\boldsymbol\varphi}}+Q-N e^{(N-1){\boldsymbol\varphi}}\right)h_{ab}\right]\,.
\eeq

Given this stress tensor we can immediately work out the boundary action that it derives from. The terms with the normal derivatives to the surface can be obtained from Gibbons-Hawking type terms, one for the metric and another for the dilaton. The subtraction terms can be easily written in terms of $d$ dimensional integrals since they are proportional to $h_{ab}$. A moment's thought makes it clear that the terms are precisely those given in \sec{s:5deff} where we have specialised to $d=4$, where we have simplified the notation by using $\frac{1}{\kappa_5^2} = \frac{\Omega_5}{8\,\pi\, G_N^{(10)}}$.


\providecommand{\href}[2]{#2}\begingroup\raggedright\endgroup

\end{document}